\documentclass[11pt]{revtex4}
\usepackage{graphicx,amsmath,amssymb}

\newcommand{\ket}[1]{\mbox{$| #1 \rangle$}}
\newcommand{\braket}[2]{\mbox{$\langle #1 | #2 \rangle$}}

\begin{document}

\title{Quantum Cryptography}
\author{Hoi-Kwong Lo and Yi Zhao}

\maketitle

\setcounter{tocdepth}{1}

\tableofcontents

\section*{Definition of the subject and its importance}
Quantum cryptography is the synthesis of quantum mechanics with the
art of code-making (cryptography). The idea was first conceived in
an unpublished manuscript written by Stephen Wiesner around 1970
\cite{Wiesner70}. However, the subject received little attention
until its resurrection by a classic paper published by Bennett and
Brassard in 1984 \cite{BB84}. The goal of quantum cryptography is to
perform tasks that are impossible or intractable with conventional
cryptography. Quantum cryptography makes use of the subtle
properties of quantum mechanics such as the quantum no-cloning
theorem and the Heisenberg uncertainty principle. Unlike
conventional cryptography, whose security is often based on unproven
computational assumptions, quantum cryptography has an important
advantage in that its security is often based on the laws of
physics. Thus far, proposed applications of quantum cryptography
include quantum key distribution (abbreviated QKD), quantum bit
commitment and quantum coin tossing. These applications have varying
degrees of success. The most successful and important
application---QKD---has been proven to be unconditionally secure.
Moreover, experimental QKD has now been performed over hundreds of
kilometers over both standard commercial telecom optical fibers and
open-air. In fact, commercial QKD systems are currently available on
the market.

On a wider context, quantum cryptography is a branch of quantum
information processing, which includes quantum computing, quantum
measurements, and quantum teleportation. Among all branches, quantum
cryptography is the branch that is closest to real-life
applications. Therefore, it can be a concrete avenue for the
demonstrations of concepts in quantum information processing. On a
more fundamental level, quantum cryptography is deeply related to
the foundations of quantum mechanics, particularly the testing of
Bell-inequalities and the detection efficiency loophole. On a
technological level, quantum cryptography is related to technologies
such as single-photon measurements and detection and single-photon
sources.

\section{Introduction}\label{se:introduction}
The best-known application of quantum cryptography is quantum key
distribution (QKD). The goal of QKD is to allow two distant
participants, traditionally called Alice and Bob, to share a long
random string of secret (commonly called the key) in the presence of
an eavesdropper, traditionally called Eve. The key can subsequently
be used to achieve a) perfectly secure communication (via
one-time-pad, see below) and b) perfectly secure authentication (via
Wigman-Carter authentication scheme), thus achieving two key goals
in cryptography.

The best-known protocol for QKD is the Bennett and Brassard protocol
(BB84) published in 1984 \cite{BB84}. The procedure of BB84 is as
follows (also shown in Table \ref{tab:bb84}).

\begin{table}\label{tab:bb84}
\center
 \caption{Procedure of BB84 protocol.}
\begin{tabular}{c|cccccccccc}
  \hline
  Alice's bit sequence & 0 & 1 & 1 & 1 & 0 & 1 & 0 & 0 & 0 & 1 \\
  \hline
  Alice's basis & + & $\times$ & + & + & $\times$ & + & $\times$ & $\times$ & + & $\times$ \\
  \hline
  Alice's photon polarization & $\rightarrow$ & $\nwarrow$ & $\uparrow$ & $\uparrow$ & $\nearrow$ & $\uparrow$ & $\nearrow$ & $\nearrow$ & $\rightarrow$ & $\nwarrow$ \\
  \hline
  Bob's basis & + & + & $\times$ & + & + & $\times$ & $\times$ & + & + & $\times$ \\
  \hline
  Bob's measured polarization & $\rightarrow$ & $\uparrow$ & $\nwarrow$ & $\uparrow$ & $\rightarrow$ & $\nearrow$ & $\nearrow$ & $\uparrow$ & $\rightarrow$ & $\nwarrow$ \\
  \hline
  Bob's sifted measured polarization & $\rightarrow$ &  &  & $\uparrow$ &  &  & $\nearrow$ &  & $\rightarrow$ & $\nwarrow$  \\
  \hline
  Bob's data sequence & 0 &  &  & 1 &  &  & 0 &  & 0 & 1 \\
  \hline
\end{tabular}
\end{table}

\begin{enumerate}
  \item Quantum communication phase
    \begin{enumerate}
      \item In BB84, Alice sends Bob a sequence of photons, each
      independently chosen from one of the four polarizations---vertical,
      horizontal, 45-degrees and 135-degrees.
      \item  For each photon, Bob randomly chooses one of the two
      measurement bases (rectilinear and diagonal) to perform a
      measurement.
      \item  Bob records his measurement bases and results. Bob
      publicly acknowledges his receipt of signals.
    \end{enumerate}
  \item Public discussion phase
    \begin{enumerate}
      \item Alice broadcasts her bases of measurements. Bob
      broadcasts his bases of measurements.
      \item Alice and Bob discard all events where they use
      different bases for a signal.
      \item To test for tampering, Alice randomly chooses a
      fraction, $p$, of all remaining events as test events. For
      those test events, she publicly broadcasts their positions
      and polarizations.
      \item Bob broadcasts the polarizations of the test events.
      \item Alice and Bob compute the error rate of the test
      events (i.e., the fraction of data for which their value
      disagree). If the computed error rate is larger than some
      prescribed threshold value, say 11\%, they abort. Otherwise,
      they proceed to the next step.
      \item Alice and Bob each convert the polarization data
      of all remaining data into a binary string called a raw key
      (by, for example, mapping a vertical or 45-degrees photon to
      ``0" and a horizontal or 135-degrees photon to ``1"). They can
      perform classical post-processing such as error correction
      and privacy amplification to generate a final key.
    \end{enumerate}
\end{enumerate}

Notice that it is important for the classical communication channel
between Alice and Bob to be authenticated. Otherwise, Eve can easily
launch a man-in-the-middle attack by disguising as Alice to Bob and
as Bob to Alice. Fortunately, authentication of an $m$-bit classical
message requires only logarithmic in m bit of an authentication key.
Therefore, QKD provides an efficient way to expand a short initial
authentication key into a long key. By repeating QKD many times, one
can get an arbitrarily long secure key.

This article is organized as follows. In Section
\ref{se:QKD_Motivation_Intro} , we will discuss the importance and
foundations of QKD; in Section III, we will discuss the principles
of different approaches to prove the unconditional security of QKD; in Section IV,
we will introduce the history and some fundamental components of QKD
implementations; in Section V, we will discuss the implementation of
BB84 protocol in detail; in Section VI, we will discuss the
proposals and implementations of other QKD protocols; in Section
VII, we will introduce a very fresh and exciting area --- quantum
hacking --- in both theory and experiments. In particular, we provide a catalogue of
existing eavesdropping attacks; in Section VIII, we will
discuss some topics other than QKD, including quantum bit
commitment, quantum coin tossing, etc.; in Section IX, we will wrap
up this article with prospectives of quantum cryptography in the
future.

\section{Quantum Key Distribution: motivation and
introduction}\label{se:QKD_Motivation_Intro}

Cryptography---the art of code-marking---has a long and
distinguished history of military and diplomatic applications,
dating back to ancient civilizations in Mesopotamia, Egypt, India
and China. Moreover, in recent years cryptography has widespread
applications in civilian applications such as electronics commerce
and electronics businesses. Each time we go on-line to access our
banking or credit card data, we should be deeply concerned with our
data security.

\subsection{Key Distribution Problem and One-time-pad}
\begin{figure}
  \includegraphics[width=10cm]{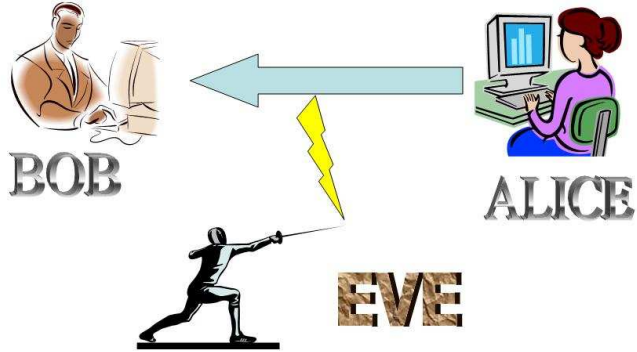}\\
  \caption{Communication in presence of an eavesdropper.}\label{Fig:AliceBobEve}
\end{figure}

Secure communication is the best-known application of cryptography.
The goal of secure communication is to allow two distant
participants, traditionally called Alice and Bob, to communicate
securely in the presence of an eavesdropper, Eve. See Figure
\ref{Fig:AliceBobEve}. A simple example of an encryption scheme is
the Caesar's cipher. Alice simply shifts each letter in a message
alphabetically by three letters. For instance, the word NOW is
mapped to QRZ, because N $\rightarrow$ O $\rightarrow$ P
$\rightarrow$ Q, O $\rightarrow$ P $\rightarrow$ Q $\rightarrow$ R
and W $\rightarrow$ X $\rightarrow$ Y $\rightarrow$ Z. According to
legends, Julius Caesar used Caesar's cipher to communicate with his
generals. An encryption by an alphabetical shift of a fixed but
arbitrary number of positions is also called a Caesar's cipher. Note
that Caesar's cipher is not that secure because an eavesdropper can
simply exhaustively try all 26 possible combinations of the key to
recover the original message.

In conventional cryptography, an unbreakable code does exist. It is
called the one-time-pad and was invented by Gilbert Vernam in 1918
\cite{OneTimePad}.
In the
one-time-pad method, a message (traditionally called the plain text)
is first converted by Alice into a binary form (a string consisting
of ``0"s and ``1"s) by a publicly known method. A key is a binary
string of the same length as the message. By combining each bit of
the message with the respective bit of the key using XOR (i.e.
addition modulo two), Alice converts the plain text into an
encrypted form (called the cipher text). i.e. for each bit $c_i =
m_i +k_i \mod 2$. Alice then transmits the cipher text to Bob via a
broadcast channel. Anyone including an eavesdropper can get a copy
of the cipher text. However, without the knowledge of the key, the
cipher text is totally random and gives no information whatsoever
about the plain text. For decryption, Bob, who shares the same key
with Alice, can perform another XOR (i.e. addition modulo two)
between each bit of the cipher text with the respective bit of the
key to recover the plain text. This is because $c_i + k_i \mod 2 =
m_i + 2k_i \mod 2 = m_i \mod 2$.

Notice that it is important not to re-use a key in a one-time-pad
scheme. Suppose the same key, k, is used for the encryption of two
messages, m1 and m2, then the cipher texts are $c_1 = m_1 + k  \mod
2$ and $c_2 = m_2+ k  \mod 2$. Then, Eve can simply take the XOR of
the two cipher texts to obtain $c_1 + c_2 \mod 2 = m_1 + m_2 + 2k
\mod 2 = m_1 + m_2 \mod 2$, thus learning non-trivial information,
namely the parity of the two messages.

The one-time-pad method is commonly used in top-secret
communication. The one-time-pad method is unbreakable, but it has a
serious drawback: it supposes that Alice and Bob initially share a
random string of secret that is as long as the message. Therefore,
the one-time-pad simply shifts the problem of secure communication
to the problem of key distribution. This is the key distribution
problem. In top-secret communication, the key distribution problem
can be solved by trusted couriers. Unfortunately, trusted couriers
can be bribed or compromised. Indeed, in conventional cryptography,
a key is a classical string consisting of ``0"s and ``1"s. In
classical physics, there is no fundamental physical principle that
can prevent an eavesdropper from copying a key during the key
distribution process.

A possible solution to the key distribution problem is public key
cryptography. However, the security of public key cryptography is
based on unproven computational assumptions. For example, the
security of standard RSA crypto-system invented by
Rivest-Shamir-Adlerman (RSA) is based on the presumed difficulty of
factoring large integers. Therefore, public key distribution is
vulnerable to unanticipated advances in hardware and algorithms. In
fact, quantum computers---computers that operate on the principles
of quantum mechanics---can break standard RSA crypto-system via the
celebrated Shor's quantum algorithm for efficient factoring
\cite{ShorAlgorithm}.

\subsection{Quantum no-cloning theorem and quantum key distribution (QKD)}

Quantum mechanics can provide a solution to the key distribution
problem. In quantum key distribution, an encryption key is generated
randomly between Alice and Bob by using non-orthogonal quantum
states. In contrast to classical physics, in quantum mechanics there
is a quantum no-cloning theorem (see below), which states that it is
fundamentally impossible for anyone including an eavesdropper to
make an additional copy of an unknown quantum state.

A big advantage of quantum cryptography is {\it forward security}.
In conventional cryptography, an eavesdropper Eve has a transcript of all communications.
Therefore, she can simply save it for many years and wait for breakthroughs such as the
discovery of a new algorithm or new hardware. Indeed, if Eve can factor large integers in
2100, she can decrypt communications sent in 2008. We remark that Canadian census data are kept
secret for 92 years on average. Therefore, factoring in the year 2100 may violate the
security requirement of our government today! And, no one in his/her sane mind can guarantee
the impossibility of efficient factoring in 2100 (except for the fact that he/she may not
live that long). In contrast, quantum cryptography guarantees forward security. Thanks to the
quantum no-cloning theorem, an eavesdropper
does {\it not} have a transcript of all quantum signals sent by Alice to Bob.

For completeness, we include the statement and the proof of the quantum
no-cloning theorem below.

\textbf{Quantum No-cloning theorem}: An unknown quantum state cannot
be copied.

(a) The case without ancilla: Given an unknown state $\ket{\alpha}$,
show that a quantum copying machine that can map $\ket{\alpha}
\ket{0} \to \ket{\alpha} \ket{\alpha} $ does not exist.

(b) The general case: Given an unknown state $\ket{\alpha}$, show
that a quantum copying machine that can map $\ket{\alpha} \ket{0}
\ket{0} \to \ket{\alpha}\ket{\alpha} \ket{u_{\alpha}}$ does not
exist.

Proof: (a) Suppose the contrary. Then, a quantum cloning machine
exists. Consider two orthogonal input states $\ket{0}$ and $\ket{1}$ respectively. We
have
$$ \ket{0}\ket{0} \to \ket{0}\ket{0}$$
and
$$ \ket{1}\ket{0} \to \ket{1}\ket{1} .$$
Consider a general input $\ket{\alpha} = a \ket{0} + b \ket{1}$.
Since a unitary transformation is linear, by linearity, we have
\begin{eqnarray}
\ket{\alpha} \ket{0} & = & ( a \ket{0} + b \ket{1} ) \ket{0}
\nonumber \\
  & \to & a \ket{0}\ket{0} + b \ket{1} \ket{1} .
  \label{wrongclone}
\end{eqnarray}
In contrast, for quantum cloning, we need:
\begin{eqnarray}
\ket{\alpha} \ket{0} & \to & ( a \ket{0} + b \ket{1} )  ( a \ket{0}
+ b \ket{1} )
\nonumber \\
  & = & a^2 \ket{0}\ket{0} + a b \ket{0} \ket{1}+
  a b \ket{1} \ket{0} + b^2 \ket{1} \ket{1} .
  \label{correctclone}
\end{eqnarray}
Clearly, if $ab \not= 0$, the two results shown in
Eqs.~(\ref{wrongclone}) and (\ref{correctclone})) are different.
Therefore, quantum cloning (without ancilla) is impossible.

(b): similar. $\Box$

More generally, for general quantum states, information gain implies
disturbance.

Theorem: (Information Gain implies disturbance) Given one state chosen from one of the two distinct
non-orthogonal states, $\ket{u}$ and $\ket{v}$ (i.e. $|
\braket{u}{v}| \not= 0$ or $1$), any operation that can learn any
information about its identity necessarily disturbs the state.

Proof: Given a system initially in state either $\ket{u}$ and
$\ket{v}$. Suppose an experimentalist applies some operation on the
system. The most general thing that she can try to do is to prepare
some ancilla in some standard state $\ket{0}$ and couple it to the
system. Therefore, we have:
\begin{equation}
 \ket{u}\ket{0} \to \ket{u} \ket{\phi_u}
\label{u0}
\end{equation}
and
\begin{equation}
 \ket{v}\ket{0} \to \ket{v} \ket{\phi_v}
\label{v0}
\end{equation}
for some states $ \ket{\phi_u}$ and $\ket{\phi_v}$.

In the end, the experimentalist lets go of the system and keeps the
ancilla. He/she may then perform a measurement on the ancilla to
learn about the initial state of the system.

Recall that quantum evolution is unitary and as such it preserves
the inner product. Now, taking the inner product between
Eqs.~(\ref{u0}) and (\ref{v0}), we get:
\begin{eqnarray}
 \braket{u}{v}\braket{0}{0} & = &\braket{u}{v}\braket{\phi_u}{\phi_v} \nonumber \\
\braket{u}{v} & = &\braket{u}{v}\braket{\phi_u}{\phi_v} \nonumber \\
\braket{u}{v} ( 1 - \braket{\phi_u}{\phi_v} ) &=& 0 \nonumber \\
  ( 1 - \braket{\phi_u}{\phi_v} ) &=& 0 \nonumber \\
  \ket{\phi_u} &= &\ket{\phi_v} ,
\label{innerproduct}
\end{eqnarray}
where in the fourth line, we have used the fact that $|
\braket{u}{v}| \not= 0$.

Now, the condition that $\ket{\phi_u} = \ket{\phi_v} $ means that
the final state of the ancilla is independent of the initial state
of the system. Therefore, a measurement on the ancilla will tell the
experimentalist nothing about the initial state of the system.
$\Box$

Therefore, any attempt by an eavesdropper to learn information about
a key in a QKD process will lead to disturbance, which can be
detected by Alice and Bob who can, for example, check the bit error
rate of a random sample of the raw transmission data.

The standard BB84 protocol for QKD was discussed in Section
\ref{se:introduction}. In the BB84 protocol, Alice prepares a
sequence of photons each randomly chosen in one of the four
polarizations---vertical, horizontal, 45-degrees and 135-degrees.
For each photon, Bob chooses one of the two polarization bases
(rectilinear or diagonal) to perform a measurement. Intuitively, the
security comes from the fact that the two polarization bases,
rectilinear and diagonal, are conjugate observables. Just like
position and momentum are conjugate observables in the standard
Heisenberg uncertainty principle, no measurement by an eavesdropper
Eve can determine the value of both observables simultaneously.
In mathematics, two conjugate observables are represented by two non-commuting
Hermitian matrices. Therefore, they cannot be simultaneously diagonalized.
This impossibility of simultaneous diagonalization implies the impossibility of
simultaneous measurements of two conjugate observables.

\subsection{Example of a simple eavesdropper attack: intercept-resend
attack}\label{se:intercept_resend}

To illustrate the security of quantum cryptography, let us consider
the simple example of an intercept-resend attack by an eavesdropper
Eve, who measures each photon in a randomly chosen basis and then
resends the resulting state to Bob. For instance, if Eve performs a
rectilinear measurement, photons prepared by Alice in the diagonal
bases will be disturbed by Eve's measurement and give random
answers. When Eve resends rectilinear photons to Bob, if Bob
performs a diagonal measurement, then he will get random answers.
Since the two bases are chosen randomly by each party, such an
intercept-resend attack will give a bit error rate of $0.5 \times
0.5 + 0.5 \times 0 = 25 \%$, which is readily detectable by Alice
and Bob. Sophisticated attacks against QKD do exist. Fortunately,
the security of QKD has now been proven. This subject will be
discussed further in Section \ref{se:security_proof}.

\subsection{Equivalence between phase and polarization encoding}

\begin{figure}
  \includegraphics[width=10cm]{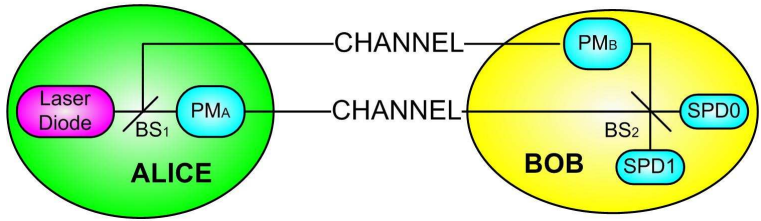}\\
  \caption{Conceptual schematic for phase-coding BB84 QKD system.
  PM: Phase Modulator; BS: Beam Splitter; SPD: Single Photon Detector.}\label{Fig:PhaseCoding}
\end{figure}

Notice that the BB84 protocol can be implemented with any two-level
quantum system (qubits).  In Section \ref{se:introduction} and the
above discussion, we have described the BB84 protocol in terms of
polarization encoding. This is just one of the many possible types
of encodings. Indeed, it should be noted that other encoding method,
particularly, phase encoding also exists. In phase encoding, a
signal consists of a superposition of two time-separated pulses,
known as the reference pulse and the signal pulse. See Figure
\ref{Fig:PhaseCoding} for an illustration of the phase encoding
scheme. The information is encoded in the relative phase between two
pulses. i.e., the four possible states used by Alice are ${ 1 /
\sqrt{2} } ( \ket{R} + \ket{S} )$ , ${ 1 / \sqrt{2} } ( \ket{R} -
\ket{S} )$, ${ 1 / \sqrt{2} } ( \ket{R} + i\ket{S} )$, ${ 1 /
\sqrt{2} } ( \ket{R} -i\ket{S} )$.

Notice that, mathematically the phase encoding scheme is equivalent
to the polarization encoding scheme. They are simply different
embodiments of the same BB84 protocol.

\section{Security Proofs}\label{se:security_proof}
``The most important question in quantum cryptography is to
determine how secure it really is." (Bennett and Brassard) Security
proofs are very important because a) they provide the foundation of
security to a QKD protocol, b) they provide a formula for the key
generation rate of a QKD protocol and c) they may even provide a
construction for the classical post-processing protocol (for error
correction and privacy amplification) that is necessary for the
generation of the final key. Without security proofs, a real-life
QKD system is incomplete because we can never be sure about how to
generate a secure key and how secure the final key really is.

\subsection{Classification of Eavesdropping attacks}

Before we discuss security proofs, let us first consider
eavesdropping attacks. Notice that there are infinitely many
eavesdropping strategies that an eavesdropper, Eve, can perform
against a QKD protocol. They can be classified as follows:

\begin{description}
  \item[Individual attacks]  In an individual attack, Eve performs
  an attack on each signal independently. The intercept-resend attack
  discussed in Section \ref{se:intercept_resend}  is an example of an individual attack.
  \item[Collective attacks:]  A more general class of attacks is
  collective attack where for each signal, Eve independently couples
  it with an ancillary quantum system, commonly called an ancilla,
   and evolves the combined signal/ancilla unitarily. She can send
   the resulting signals to Bob, but keep all ancillas herself. Unlike
   the case of individual attacks, Eve postpones her choice of
   measurement. Only after hearing the public discussion between Alice
and Bob, does Eve decide on what measurement to perform on her
ancilla to extract information about the final key.
  \item[Joint attacks:] The most general class of attacks is joint
  attack. In a joint attack, instead of interacting with each signal
   independently, Eve treats all the signals as a single quantum system.
    She then couples the signal system with her ancilla and evolves the
    combined signal and ancilla system unitarily. She hears the public
    discussion between Alice and Bob before deciding on which measurement
    to perform on her ancilla.
\end{description}

Proving the security of QKD against the most general attack was a
very hard problem. It took more than 10 years, but the unconditional
security of QKD was finally established in several papers in the
1990s. One approach by Mayers \cite{Security:Mayers} was to prove
the security of the BB84 directly. A simpler approach by Lo and Chau
\cite{Security:Lo_Science}, made use of the idea of entanglement
distillation by Bennett, DiVincenzo, Smolin and Wootters (BDSW)
\cite{BDSW1996} and quantum privacy amplification by Deutsch et al.
\cite{Security:Deutsch_PRL96} to solve the security of an
entanglement-based QKD protocol. The two approaches have been
unified by the work of Shor and Preskill
\cite{Security:ShorPreskill}, who provided a simple proof of
security of BB84 using entanglement distillation idea. Other early
security proofs of QKD include Biham, Boyer, Boykin, Mor, and
Roychowdhury \cite{Security:Biham_STOC00}, and Ben-Or
\cite{Security:Ben-or}.

\subsection{Approaches to security proofs}

There are several approaches to security proof. We will discuss them
one by one.

\begin{enumerate}
  \item Entanglement distillation

  Entanglement distillation protocol (EDP) provides a simple approach
   to security proof
   \cite{Security:Deutsch_PRL96,Security:Lo_Science,Security:ShorPreskill}.
   The basic insight is that entanglement is a sufficient (but not
   necessary) condition for a secure key. Consider the noiseless case
   first. Suppose two distant parties, Alice and Bob, share a maximally
   entangled state of the form $\ket{\phi}_{AB}= { 1 / \sqrt{2} } (
   \ket{00}_{AB} + \ket{11}_{AB} )$. If each of Alice and Bob measure
   their systems, then they will both get "0"s or "1"s, which is a shared
   random key. Moreover, if we consider the combined system of the three
    parties---Alice, Bob and an eavesdropper, Eve, we can use a pure-state
     description (the ``Church of Larger Hilbert space") and consider a
     pure state $\ket{\psi}_{ABE}$. In this case, the von Neumann
     entropy \cite{von_Neumann_entropy}
     of Eve $S(\rho_E) = S(\rho_{AB}) =0$.
    This means that Eve has absolutely no information on the final key.
   This is the consequence of the standard Holevo's theorem. See, like, \cite{Holevo_Theorem_Wiki}.

    In the noisy case, Alice and Bob may share $N$ pairs of qubits, which
     are a noisy version of $N$ maximally entangled states. Now, using the
     idea of entanglement distillation protocol (EDP) discussed in BDSW \cite{BDSW1996},
     Alice and Bob may apply local operations and classical communications
     (LOCCs) to distill from the $N$ noisy pairs a smaller number, say
     $M$
     almost perfect pairs i.e., a state close to $\ket{\phi}_{AB}^M$. Once
     such a EDP has been performed, Alice and Bob can measure their
     respective system to generate an $M$-bit final key.

     One may ask: how can Alice and Bob be sure that their EDP will be
     successful? Whether an EDP will be successful or not depends on the
     initial state shared by Alice and Bob. In the above, we have skipped
     the discussion about the verification step. In practice, Alice and Bob
     can never be sure what initial state they possess. Therefore, it is
     useful for them to add a verification step. By, for example, randomly
      testing a fraction of their pairs, they have a pretty good idea about
      the properties (e.g., the bit-flip and phase error rates) of their remaining pairs and are pretty confident that
      their EDP will be successful.

     The above description of EDP is for a quantum-computing protocol
      where we assume that Alice and Bob can perform local quantum
      computations. In practice, Alice and Bob do not have large-scale
      quantum computers at their disposal. Shor and Preskill made the
       important observation that the security proof of the
       standard
       BB84 protocol can be reduced to that of an EDP-based QKD protocol \cite{Security:Deutsch_PRL96,Security:Lo_Science}.
         The Shor-Preskill proof \cite{Security:ShorPreskill} makes use of
         the Calderbank-Shor-Steane (CSS) code, which has the advantage of
         decoupling the quantum error correction procedure into two parts:
         bit-flip and phase error correction. They can go on to show that
          bit-flip error correction corresponds to standard error
          correction and phase error correction corresponds to privacy
           amplification (by random hashing).

  \item Communication complexity/quantum memory.

  The communication complexity/quantum memory approach to security proof
  was proposed by Ben-Or \cite{Security:Ben-or} and subsequently by
  Renner and Koenig \cite{Security:RennerKoenig}. See also
  \cite{Security:Renner}. They provide a formula for secure key generation
  rate in terms of an eavesdropper's quantum knowledge on the raw
  key:  Let $Z$ be a random variable with range $\mathcal{Z}$, let $\boldsymbol{\rho}$ be a
  random state, and let $F$ be a two-universal{} function on $\mathcal{Z}$ with range
  $\mathcal{S} = \{0,1\}^s$ which is independent of $Z$ and $\boldsymbol{\rho}$.
  Then \cite{Security:RennerKoenig}
  \[
    d(F(Z)|\{F\} \otimes \boldsymbol{\rho})
  \leq
     \frac{1}{2} 2^{-\frac{1}{2} (S_2([\{Z\} \otimes \boldsymbol{\rho}])
     - S_0([\boldsymbol{\rho}]) - s)} \ .
  \]

Incidentally, the quantum de Finnetti's theorem
\cite{Security:SymmetryIndependenceRenner} is often useful for
simplifying security proofs of this type.

  \item Twisted state approach.

  What is a necessary and sufficient condition for secure key generation?
  From the entanglement distillation approach, we know that entanglement
  distillation a sufficient condition for secure key generation. For some
  time, it was hoped that entanglement distillation is also a necessary
  condition for secure key generation. However, such an idea was proven
   to be wrong in \cite{Security:HHHO,Security:ZeroQuantumCapacity}, where it was found that
      a necessary and sufficient condition is the distillation
 of a private state, rather than a maximally entangled state. A private state
  is a ``twisted" version of a maximally entangled state. They
  proved the following theorem in \cite{Security:HHHO}: a state is private in the above sense
iff it is of the following form
\begin{equation}\label{eq:pstate}
\gamma_m=U|\psi_{2^m}^+\rangle_{AB}\langle\psi_{2^m}^+|\otimes
\varrho_{A'B'}U^\dagger
\end{equation}
where $\ket{\psi_{d}}=\sum_{i=1}^d \ket{ii}$ and $\varrho_{A'B'}$ is
an arbitrary state on $A'$,$B'$. $U$ is an arbitrary unitary
controlled in the computational basis
\begin{equation}\label{eq:u}
U=\sum_{i,j=1}^{2^m} |ij\rangle_{AB}\langle ij|\otimes U_{ij}^{A'B'}.
\end{equation}

The operation (\ref{eq:u}) will be called ``twisting" (note that
only $U_{ii}^{A'B'}$  matter here, yet it will be useful to consider
general twisting later).

{\it Proof. (copied from \cite{Security:HHHO})} The authors of
\cite{Security:HHHO} proved for $m=1$ (for higher $m$, the proof is
analogous). Start with an arbitrary state held by Alice and Bob,
$\rho_{AA'BB'}$, and include its purification to write the total
state in the decomposition
\begin{eqnarray} && \Psi_{ABA'B',E}=
a|00\rangle_{AB}|\Psi_{00}\rangle_{A'B'E} +
b|01\rangle_{AB}|\Psi_{01}\rangle_{A'B'E} \nonumber\\
&&+c|10\rangle_{AB}|\Psi_{10}\rangle_{A'B'E} +
d|11\rangle_{AB}|\Psi_{11}\rangle_{A'B'E} \end{eqnarray} with the
states $\ket{ij}$ on $AB$ and $\Psi_{ij}$ on $A'B'E$. Since the key
is unbiased and perfectly correlated, we must have
 $b=c=0$ and $|a|^2=|d|^2=1/2$.
Depending on whether the key is $\ket{00}$ or $\ket{11}$, Eve will
hold the states
\begin{equation}
\varrho_{0}=Tr_{A'B'}|\Psi_{00}\rangle \langle \Psi_{00}|,\quad
\varrho_{1}=Tr_{A'B'}|\Psi_{11}\rangle \langle \Psi_{11}|
\end{equation}
Perfect security requires $\varrho_0=\varrho_1$. Thus there exists
unitaries $U_{00}$ and $U_{11}$ on $A'B'$ such that \begin{eqnarray}
|\Psi_{00}\rangle&=&\sum_{i}\sqrt{p_{i}}
|U_0\phi_{i}^{A'B'}\rangle|\varphi_{i}^{E}\rangle \nonumber\\
|\Psi_{11}\rangle&=&\sum_{i}\sqrt{p_{i}}
|U_1\phi_{i}^{A'B'}\rangle|\varphi_{i}^{E}\rangle. \end{eqnarray}
After tracing out $E$, we will thus get a state of the form Eq.
(\ref{eq:pstate}), where
$\varrho_{A'B'}=\sum_ip_i|\phi_{i}\>\langle\phi_i|\Box$.

  The main new ingredient of the above theorem is the introduction of a ``shield"
  part to Alice and Bob's system. That is, in addition to the systems $A$ and
  $B$ used by Alice and Bob for key generation, we assume that Alice and Bob
  also hold some ancillary systems, $A'$ and $B'$, often called the shield part.
  Since we assume that Eve has no access to the shield part, Eve is further
  limited in her ability to eavesdrop. Therefore, Alice and Bob can derive a
  higher key generation rate than the case when Eve does have access to the
  shield part.

  An upshot is that even a bound entangled state can give a secure key. A bound
  state is one whose formation (via local operations and classical communications,
  LOCCs) requires entanglement, but which does not give any distillable entanglement.
  In other words, even though no entanglement can be distilled from a bound entangled
  state, private states (a twisted version of entangled states) {\it can} be
  distilled from a bound entangled state.

  In summary, secure key generation is a more general theory than entanglement distillation.

  \item Complementary principle

  Another approach to security proof is to use the complementary principle
  of quantum mechanics. Such an approach is interesting because it shows
  the deep connection between the foundations of quantum mechanics and the
  security of QKD. In fact, both Mayers' proof \cite{Security:Mayers} and
  Biham, Boyer, Boykin, Mor, and
Roychowdhury's proof \cite{Security:Biham_STOC00} make
  use of this complementary principle. A clear and rigorous discussion of
  the complementary principle approach to security proof has recently been
  achieved by Koashi \cite{Security:Koashi_Complementary}.

  The key insight of Koashi's proof is that Alice and Bob's ability to generate a
  random secure key in the $Z$-basis (by a measurement of the Pauli spin matrix
  $\sigma_Z$)  is equivalent to the ability for Bob to help Alice prepare an eigenstate
  in the complementary, i.e., $X$-basis ($\sigma_X$), with their help of the shield. The
  intuition is that an $X$-basis eigenstate, for example,
  $\ket{+}_A = { 1 \over \sqrt{2} } (\ket{0}_A + \ket{1}_A )$, when measured along the
  $Z$-basis, gives a random answer.

  \item Other ideas for security proofs

  Here we discuss two other ideas for security proofs, namely, a) device-independent security proofs
  and b) security from the causality constraint. Unfortunately, these ideas are still
  very much under development and so far a complete version of a proof of unconditional
  security of QKD based on
  these ideas with a finite key rate is still missing.

  Let us start with a) device-independent security proofs
  So far we have assumed that Alice and Bob know what their devices are doing
  exactly. In practice, Alice and Bob may not know their devices for sure. Recently,
  there has been much interest in the idea of device-independent security proofs. In other
  words, how to prove security when
  Alice and Bob's
  devices cannot be trusted. See, for example, \cite{Security:DeviceIndependent}. The idea is to look only at the input and output variables. A handwaving argument goes as follows.
  Using their probability distribution, if one can demonstrate the violation of some Bell
  inequalities, then one cannot explain the data by a separable system.
  How to develop such a handwaving argument into a full proof of unconditional security is
  an important question.

  The second idea b) security from the causality constraint is even more ambitious.
  The question that it tries to address is the following.
  How can one prove security when even quantum mechanics is wrong?
  In \cite{Security:Causality}) and references cited therein, it was suggested
  that perhaps a more general physical
  principle such as the no-signaling requirement for space-like observables
  could be used to prove the security of QKD.

\end{enumerate}

\subsection{Classical Post-Processing Protocols}

As noted in Section \ref{se:introduction}, after the quantum
communication phase, Alice and Bob then proceed with the classical
communication phase. In order to generate a secure key, Alice and
Bob have to know what classical post-processing protocol to apply to
the raw quantum data. This is a highly non-trivial question. Indeed,
\emph{a priori}, given a particular procedure for classical
post-processing, it is very hard to know whether it will give a
secure key or what secure key will be generated. In fact, it is
sometimes said that in QKD, the optical part is easy, the
electronics part is harder, but the hardest part is the classical
post-processing protocol. Fortunately, security proofs often give
Alice and Bob clear ideas on what classical post-processing protocol
to use. This highlights the importance for QKD practitioners to
study the security proofs of QKD.

Briefly stated, the classical post-processing protocol often
consists of a) test for tampering and b) key generation. In a) test
for tampering, Alice and Bob may randomly choose a fraction of the
signals for testing. For example, by broadcasting the polarizations
of those signals, they can work out the bit error rate of the test
signals. Since the test signals are randomly chosen, they have high
confidence on the bit error rate of the remaining signals. If the
bit error rate of the tested signal is higher than a prescribed
threshold value, Alice and Bob abort. On the other hand, if the bit
error rate is lower than or equal to the prescribed value, they
proceed with the key generation step with the remaining signals.
They first convert their polarization data into binary strings, the
raw keys, in a prescribed manner. For example, they can map a
vertical or 45-degrees photon to ``0" and a horizontal or
135-degrees photon to ``1". As a result, Alice has a binary string
$x$ and Bob has a binary string $y$. However, two problems remain.
First, Alice's string may differ from Bob's string. Second, since
the bit error rate is non-zero, Eve has some information about
Alice's and Bob's string. The key generation step may be divided
into the following stages:

\begin{enumerate}
  \item Classical pre-processing

  This is an optional step. Classical pre-processing has the
  advantage of achieving a higher key generation rate and tolerating
  a higher bit error rate \cite{Security:GottesmanLo,Security:RennerGisinKraus,Security:Chau}.

  Alice and Bob may pre-process their data by either a) some type of
  error detection algorithm or b) some random process. An example of
  an error detection algorithm is a B-step \cite{Security:GottesmanLo}, where
  Alice randomly permutes all her bits and broadcasts the parity of
  each adjacent pair. In other words, starting from
  $\vec{x}= (x_1, x_2, \cdots x_{2N-1}, x_{2N} )$, Alice broadcasts a
  string $\vec{x_1} = ( x_{\sigma (1)} + x_{\sigma (2)} \mod 2 , x_{\sigma(3)} + x_{\sigma(4)} \mod2  , \cdots , x_{\sigma(2N-1)} + x_{\sigma(2N)} \mod 2 )$,
  where $\sigma$ is a random permutation chosen by Alice. Moreover, Alice
  informs Bob which random permutation, $\sigma$, she has chosen. Similarly,
  starting from $\vec{y}= (y_1, y_2, \cdots y_{2N-1}, y_{2N} )$, Bob randomly
  permutes all his bits using the same $\sigma$ and broadcasts the parity bit of
  all adjacent pairs. I.e.. Bob broadcasts
  $\vec{y_1} = ( y_{\sigma(1)} + y_{\sigma(2)} \mod 2 , y_{\sigma(3)} + y_{\sigma(4)} \mod2  , \cdots , y_{\sigma(2N-1)} + y_{\sigma(2N)} \mod 2 )$.
  For each pair of bits, Alice and Bob keep the first bit iff their parities
  of the pair agree. For instance, if
  $x_{\sigma (2k -1) } + x_{\sigma (2k)} \mod 2  = y_{\sigma (2k-1) } + y_{\sigma (2k) } \mod 2 $, then Alice keeps
  $x_{\sigma (2k-1) }$ and Bob keeps $y_{\sigma (2k-1) }$ as their new key bit. Otherwise,
  they drop the pair $(x_{\sigma (2k-1) }, x_{\sigma (2k) }) $
  and $(y_{\sigma (2k-1) } x_{\sigma (2k) })$ completely.

  Notice that the above protocol is an error detection protocol. To see
  this, let us regard the case where $x_i \not= y_i $ as an error during
  the quantum transmission stage. Suppose that for each bit, $i$, the event
  $x_i \not= y_i $ occurs with an independent probability $p$. For each $k$,
  the B step throws away the cases where a single error has occurred for the
  two locations $\sigma (2k-1) $ and $\sigma (2k)$ and keeps the cases when
  either no error or two errors has occurred. As a result, the error probability
  after the B-step is reduced from $O(p)$ to $O(p^2)$. The random permutation
  of all the bit locations ensures that the error model can be well described
  by an independent identical distribution (i.i.d.).

  An example of a random process is an adding noise protocol \cite{Security:RennerGisinKraus}
  where, for each bit $x_i$, Alice randomly and independently chooses to keep it
  unchanged or flip it with probabilities, $1-q$ and $q$ respectively, where
  the probability $q$ is publicly known.

  \item Error correction

  Owing to noises in the quantum channel, Alice and Bob's raw keys, $x$
  and $y$, may be different. Therefore, it is necessary for them to reconcile
  their keys. One simple way of key reconciliation is forward key reconciliation,
  whose goal is for Alice to keep the same key $x$ and Bob to change his key from
  $y$ to $x$. Forward key reconciliation can be done by either standard error
  correcting codes such as low-density-parity-check (LDPC) codes or specialized
  (one-way or interactive) protocols such as Cascade
  protocol \cite{Security:ErrorCorrectionEfficiency}.

  \item Privacy amplification

  To remove any residual information Eve may have about the key, Alice and Bob
  may apply some algorithm to compress their partially secure key into a shorter
  one that is almost perfectly secure. This is called privacy amplification. Random
  hashing and a class of two-universal hash functions are often suitable for privacy
  amplification. See for example \cite{Security:GeneralPrivacyAmplification}
  and
  \cite{Security:Renner} for discussion.

\end{enumerate}

\subsection{Composability}

A key generated in QKD is seldom used in isolation. Indeed, one may
concatenate a QKD process many times, using a small part of the key
for authentication each time and the remaining key for other
purposes such as encryption. It is important to show that using QKD
as a sub-routine in a complicated cryptographic process does not
create new security problems. This issue is called the composability
of QKD and, fortunately, has been solved in
\cite{Security:Composability}.

Composability of QKD is not only of academic interest. It allows us
to refine our definition of security
\cite{Security:Composability,Security:LockingAccessibleInformation}
and directly impacts on the parameters used in the classical
post-processing protocol.

\subsection{Security proofs of practical QKD systems}

As will be discussed in Section \ref{se:ExpComponent}, practical QKD
systems suffer from real-life imperfections. Proving the security of
QKD with practical systems is a hard problem. Fortunately, this has
been done with semi-realistic models by Inamori, L\"{u}tkenhaus and Mayers \cite{ILM} and
in a more general setting by Gottesman, Lo, L\"{u}tkenhaus, and Preskill
\cite{GLLP}.

\section{Experimental Fundamentals}\label{se:ExpComponent}
Quantum cryptography can ensure the secure communication between two
or more legitimate parties. It is more than a beautiful idea.
Conceptually, it is of great importance in the understandings of
both information and quantum mechanics. Practically, it can provide
an ultimate solution for confidential communications, thus making
everyone's life easier.

By implementing the quantum crypto-system in the real life, we can
test it, analyze it, understand it, verify it, and even try to break
it. Experimental quantum key
distribution (QKD) has been performed since about 1989 and great progress has been
made. Now,
you can even buy QKD systems on the market.

A typical QKD set-up includes of three standard parts: the source
(Alice), the channel, and the detection system (Bob).

\subsection{A brief history}
\subsubsection{The first experiment}
The proposal of BB84 \cite{BB84} protocol seemed to be simple.
However, it took another five years before it was first
experimentally demonstrated by Bennett, Bessette, Brassard, Salvail,
and Smolin in 1989 \cite{EQC}. This first demonstration was based on
polarization coding. Heavily attenuated laser pulses instead of
single photons were used as quantum signals, which were transmitted
over 30cm open
air at a repetition rate of 10Hz. 


\subsubsection{From centimeter to kilometer}
30cm is not that appealing for practical communications. This short
distance is largely due to the difficulty of optical alignment in
free space. Switching the channel from open air to optical fiber is
a natural choice. In 1993, Townsend, Rarity, and Tapster
demonstrated the feasibility of phase-coding fiber-based QKD over
10km telecom fiber \cite{Townsend:EL1993} and Muller, Breguet, and
Gisin demonstrated the feasibility of polarization-coding
fiber-based QKD over 1.1km telecom fiber \cite{Muller:EPL1993}.
(Also, Jacobs and Franson demonstrated both free-space
\cite{ExpQKD:FreeSpace_OL96} and fiber-based QKD
\cite{ExpQKD:Fiber_AO94}.) These are both feasibility demonstrations
by means that neither of them applied random basis choosing at Bob's
side. Townsend's demonstration seemed to be more promising than
Muller's due to the following reasons.
\begin{enumerate}
    \item The polarization dispersion in fibers is highly
    unpredictable and unstable. Therefore polarization coding
    requires much more controlling in the fiber than phase coding.
    In fiber-based QKD implementations, phase coding is in general
    more preferred than polarization coding.
    \item Townsend et al. used 1310ns laser as the source, while
    Muller et al. used 800ns laser as the source.
    1310nm is the second window wavelength of  telecom fibers
    (the first window wavelength is 1550nm). The
    absorption coefficient of standard telecom fiber at 1310nm is 0.35dB/km, comparing to
    3dB/km at 800nm. Therefore the fiber is more transparent to Townsend et al.'s set-up.
\end{enumerate}

P. D. Townsend demonstrated QKD with Bob's random basis selection in
1994 \cite{Townsend:EL1994}. It was phase-coding and was over 10km
fiber. The source repetition rate was 105MHz (which is quite high
even by today's standard) but the phase modulation rate was 1.05MHz.
This mismatch brought a question mark on its security.

\subsubsection{Getting out of the lab}
It is crucial to test QKD technique in the field deployed fiber.
Muller, Zbinden, and Gisin successfully demonstrated the first QKD
experiment outside the labs with polarization coding in 1995
\cite{Muller:Nature1995,Muller:EPL1996}. This demonstration was
performed over 23km installed optical fiber under Lake Geneva.
(Being under water, quantum communication in the optical fiber
suffered less noise.)



There is less control over the field deployed fiber than fiber in
the labs. Therefore its stabilization becomes challenging. To solve
this problem, A. Muller et al. designed the ``plug \& play''
structure in 1997 \cite{Muller:APL97}. A first experiment of this
scheme was demonstrated  by H. Zbinden et al. in the same year
\cite{Zbinden:EL97}. Stucki, Gisin, Guinnard, Robordy, and Zbinden
later demonstrated a simplified version of the ``plug \& play''
scheme under Lake Geneva over 67 km telecom fiber in 2002
\cite{Stucki:NJP2002}.

\subsubsection{With a coherent laser source}

The original BB84 \cite{BB84} proposal required a single photon
source. However, most QKD implementations are based on faint lasers
due to the great challenge to build the perfect single photon
sources. In 2000, the security of coherent laser based QKD systems
was analyzed first against individual attacks in 2000
\cite{Lutkenhaus:PRA00}. Finally, the unconditional security of
coherent laser based QKD systems was proven in 2001 \cite{ILM} and
in a more general setting in 2002 \cite{GLLP}. Gobby, Yuan, and
Shields demonstrated an experiment based on \cite{Lutkenhaus:PRA00}
in 2005 \cite{ExpQKD:50km_Secure} (Note that this work was claimed
to be unconditionally secure. However, due to the limit of
\cite{Lutkenhaus:PRA00}, this is only true against individual
attacks rather than the most general attack).

The security analysis in \cite{ILM,GLLP} will
severely limit the performance of unconditonally secure QKD systems. Fortunately, since 2003
the decoy state method has been proposed \cite{Decoy:Hwang,Decoy:LoISIT,Decoy:LoPRL,Decoy:Practical,Decoy:WangPRL,Decoy:WangPRA} by Hwang and extensively analyzed by our group at
the University of Toronto and by Wang.
The first experimental demonstration of decoy state QKD was reported
by us in 2006 \cite{Decoy:ZhaoPRL} over 15 km telecom
fiber and later over 60 km telecom fiber \cite{Decoy:ZhaoISIT}.
Subsequently, decoy state QKD was further demonstrated by
several other groups
\cite{Decoy:144km,Decoy:LATES,Decoy:YuanAPL,Decoy:PanPRL,Decoy:130km}.
The readers may refer to Section \ref{se:protocol_decoy} for details
of decoy state protocols.

  \subsection{Sources}
  \begin{description}
    \item[Single Photon Sources] are demanded by the original BB84
    \cite{BB84} proposal. Suggested by its name, the single photon
    sources are expected to generate exactly one photon on demand.
    The bottom line for a single photon source is that no more than
    one photon can be generated at one time.
    It is very hard to build a perfect single photon source (i.e., no multi-photon
    production). Despite
    tremendous effort made by many groups, perfect  single photon
    source is still far from practical. Fortunately, the proposal and
    implementation of decoy state QKD (see Section
    \ref{se:protocol_decoy}) make it unnecessary to use single
    photon sources in QKD.

    \item[Parametric Down-conversion (PDC) Sources] are often used as the
    entanglement source. Its principle is that a high energy
    ($\sim$400nm) photon propagates through a highly non-linear
    crystal (usually BBO), producing two entangled photons
    with frequency halved. PDC sources are usually used for
    entanglement-based QKD systems (eg. Ekert91 \cite{Ekert91}
    protocol).

    PDC sources are also used as ``triggered single photon
    sources'', in which Alice possesses a PDC source and monitor one
    arm of its outputs. In case that Alice sees a detection, she
    knows that there is one photon emitted from the other arm.
    Experimental demonstration of QKD with PDC sources is reported
    in \cite{Decoy:ExpPDC_Wang}.

    \item[Attenuated Laser Sources] are the most commonly used
    sources in QKD experiments. They are essentially the same as
    the laser sources used in classical optical communication except
    for that heavy attenuation is applied on them. They are simple and reliable, and
    they can reach Gigahertz with little challenge.
    In BB84 system and differential-phase-shift-keying (DPSK) system (to be discussed in
    Subsection~\ref{se:protocol_DPSK}), the laser
    source is usually attenuated to below 1 photon per pulse. In
    Gaussian-modulated coherent-state (GMCS)
    system (to be discussed in Subsection~\ref{se:protocol_GMCS}),
    the laser source is usually attenuated to around
    100 photons per pulse.

    Attenuated laser sources used to be considered to be non-ideal for BB84
    systems as they always have non-negligible probability of
    emitting multi-photon pulses regardless how heavily they are
    attenuated. However, the discovery and implementation of decoy
    method
    \cite{Decoy:Hwang,Decoy:LoISIT,Decoy:LoPRL,Decoy:Practical,Decoy:WangPRL,Decoy:WangPRA,Decoy:ZhaoPRL,Decoy:ZhaoISIT,Decoy:144km,Decoy:PanPRL,Decoy:LATES,Decoy:YuanAPL}
    made coherent laser source much more appealing. With decoy method, it
    is possible to make BB84 system with laser source secure without
    significant losses on the performance.

    \end{description}

  \subsection{Channels}
  \begin{description}
    \item[Standard Optical Single-Mode Fiber (SMF)] is the most popular choice for now. It can
    connect two arbitrary points, and can easily be extended to
    networks. Moreover, it is deployed in most developed urban
    areas.

    SMF has two ``window wavelengths'': one is 1310nm and the other
    is 1550nm. The absorptions at these two wavelengths are
    particularly low ($\sim0.35$dB/km at 1310nm, and $\sim0.21$dB/km at
    1550nm). Nowadays most fiber-based QKD implementations use
    1550nm photons as information carriers.

    The main disadvantage of optical fiber is its birefringence. The
    strong polarization dispersion made it hard to implement
    polarization-coding system. Also it has strong spectral
    dispersion, which affects the high speed (10+ GHz) QKD systems heavily
    \cite{DPSK:200km} as the pulses are broadened and overlap with
    each other. For this reason, the loss in fibers (0.21dB/km at 1550nm) puts
    an limit on the longest distance that a fiber-based QKD system
    can reach.
    \item[Free Space] is receiving more and more attention recently.
    It is ideal for the polarization coding. There is
    negligible dispersion on the polarization and the frequency. However,
    the alignment of optical beams can be challenging for long
    distances, particularly due to the atmospheric turbulence.
    Notice that open-air QKD requires a direct line of sight between Alice and Bob
    (unless some forms of mirrors are used).
    Buildings and mountains are serious obstacles
    for open-air QKD systems.

    The greatest motivation for open-air QKD scheme is the hope for
    ground-to-satellite \cite{Proposal:GroundSatellite}
    and satellite-to-satellite quantum communication.
    As there is negligible optical absorption in the outer space, we may
    be able to achieve inter-continental quantum communication with
    free-space QKD.
    \end{description}

  \subsection{Detection systems}\label{se:detection}
  \begin{description}
    \item[InGaAs-APD Single Photon Detectors] are the most popular
    type of single-photon detectors in fiber-based QKD and they are
    commercially available.
    InGaAs-APD Single Photon Detectors utilize the avalanche effect of
    semiconductor diodes. A strong biased voltage is applied on the InGaAs
    diode. The incident photon will trigger the avalanche effect, generating
    a detectable voltage pulse. The narrow band gap of InGaAs made it possible to
    detect photons at telecom wavelengths (1550nm or 1310 nm).

    InGaAs APD based single photon detectors have simple structure
    and commerically packaged. They are easy to calibrate and
    operate. The reliability of InGaAs APD is relatively high. They
    normally work at $-50^\circ$C to $-110^\circ$C to lower the dark
    count rate. This temperature can be easily achieved by
    thermal-electric coolers.
   The detection efficiency of InGaAs-APD
    based single photon detectors is usually $\sim10\%$
    \cite{InGaAsAPD:Stucki01}.

    In single photon detectors, a key parameter (besides detection efficiency)
    is the dark count rate.
    The dark count is the event that the detector generates a detection click while
    no actual photon hits it (i.e. ``false alarm''). The dark count rate of InGaAs
    single photon detector is relatively high ($~ 10^{-5}$ per gate. The concept of
    gating will be introduced below.) even if it is cooled.

    The after-pulse effect is that the dark count rate of the detector increases
    for a time period after a successful detection. This effect is serious for InGaAs
    single photon detectors. Therefore the blank circuit
    is often introduced to reduce this effect. The mechanism of the blank circuit is that
    the detector is set to be deactivated for a time period, which is called the ``dead time'',
    after a detection event. The dead time should be set to long enough so that
    when the detector is re-activated, the after-pulse effect is negligible.
     The dead time for InGaAs single photon
    detector is typically in the order of microseconds \cite{InGaAsAPD:Stucki01}.
    The long after pulse effect, together with the large timing jitter limits the
    InGaAs-APD based single photon detectors
    to work no faster than several megahertz. Moreover, the blank
    circuit reduces the detection efficiency of InGaAs-APD based
    single photon detectors.

    An additional method to reduce the dark count rate is to apply
    the gating mode. i.e., the detectors are only activated when the
    photons are expected to hit them. Gating mode reduces the dark count rate
    by several orders and is thus used in most
    InGaAs-APD single photon detectors. However, it may open up a security
    loophole \cite{Qi:TimeShift,Zhao:TimeShift}.

    There is a trade-off between the detection efficiency and the dark count
    rate. As the biased voltage on an APD increases, both the
    detection efficiency and the dark count rate increase.

    Recently, it has been reported that, by gating an InGaAs detector in a sinusoidal manner,
    it is possible to reduce the dead time and operate a QKD system at 500MHz. See
    \cite{DPSK:500MHz}.
    This result seems to be an important development
    which could make InGaAs detectors competitive with newer single photon
    detector technologies such as SSPDs (to be introduced below).

    \item[Si-APD Single Photon Detectors] are ideal for detection of
    visible photons (say ~800nm). They have negligible dark count rate and can
    work at room temperature. They are very compact in size. More
    importantly, they have high detection efficiency ($>60\%$) and
    can work at gigahertz. These
    detectors are ideal for free space QKD systems. However, the
    band gap of silicon is too large to detect photons at telecom wavelength (1550nm or 1310nm),
    and the strong attenuation of telecom fiber on visible
    wavelengths makes it impractical to use visible photons in
    long distance fiber-based QKD systems.

    \item[Parametric Up-conversion Single Photon Detectors] try to
    use Si-APD to detect telecom wavelength photons. It uses
    periodically poled lithium niobate (PPLN) waveguide and a pumping
    light to up-convert the incoming telecom frequency
    photons into visible frequency, and uses Si-APD to detect these
    visible photons. The high speed and low timing-jitter of Si-APDs make it possible to
    perform GHz QKD on fiber-based system with up-conversion single photon
    detectors \cite{ExpQKD:GHz,DPSK:100km}.

    The efficiency of up-conversion detectors is similar to that of
    InGaAs APD single photon detectors. There is also a trade-off
    between the detection efficiency and the dark count rate. When
    increasing the power of the pumping light, the conversion
    efficiency will increase, improving the detection efficiency.
    Meanwhile, more pumping photons and up-converted pumping photons
    (with frequency doubled) will pass through the filter and enter
    the Si-APD, thus increasing the dark count rate \cite{DPSK:100km}.

    \item[Transiting-edge Sensor (TES)] is based on critical state
    superconductor rather than semiconductor APDs. It uses squared
    superconductor (typically tungsten) thin film as ``calorimeter'' to measure the
    electron temperature. A biased voltage is applied on the thin film
    to keep it in critical state. Once one or more photons are absorbed by the
    sensor, the electron temperature will change, leading to a change of
    the current. This current change can be
    detected by a superconductive quantum-interference device (SQUID)
    array \cite{ExpQKD:TES_50km}.

    The TES single photon detectors can achieve very high detection
    efficiency (up to 89\%) at telecom wavelength. The dark count rate is negligible.
     Moreover, TES detectors can resolve photon numbers. This is because the
    electron temperature change is proportional to the number of
    photons that have been absorbed.

    The thermal nature of TES detectors limits their counting rates.
    Once some photons were absorbed by the sensor, it would take a
    few microseconds before the heat is dissipated to the substrate.
    This long relaxation time limits the counting rate of TES
    detector to no more than a few megahertz \cite{ExpQKD:TES_50km}.
    This is a major drawback of TES.

    The bandwidth of TES detector is extremely wide. The detector is
    sensitive to all the wavelengths. Even the black body radiation
    from the fiber or the environment can trigger the detection
    event, thus increasing the dark count rate. To reduce the dark counts
    caused by other wavelengths, a spectral filter is necessary.
    However, this will increase the internal loss and thus
    reduce the detection efficiency.

    One of the greatest disadvantage of TES detector is its working
    temperature: 100mK. This temperature probably requires
    complicated cooling devices \cite{ExpQKD:TES_50km}.

    \item[Superconductive Single Photon Detectors (SSPDs)] also use
    superconductor thin film to detect incoming photons. However, instead of using a piece of plain thin
    film, a pattern of zigzag superconductor (typically NbN) wire is formed.
    The superconductive wire is set to critical state by applying
    critical current through it. Once a photon hits the wire, it
    heats a spot on the wire and makes the spot over-critical
    (i.e., non-superconductive.). As the current is the same as
    before, the current density in the areas around this hot-spot
    increases, thus making these areas non-superconductive. As a result,
    a section of the wire becomes non-superconductive, and
    a voltage spike can be observed as the current is kept constant \cite{DPSK:200km}.

    The SSPD can achieve very high (up to 10GHz) counting rate. This is because
    the superconductor wire used in SSPD can dissipate the heat in
    tens of picoseconds. It also has very low dark count rate (around
    10Hz) due to the superconductive nature. The SSPDs should be
    able to resolve the incident photon number in principle.
    However, photon number resolving SSPDs have never been reported
    yet.

    The efficiency of SSPD is lower than that of TES. This is because only part ($\sim50\%$) of
    the sensing area is covered by the wire. The fabrication of such
    complicated zigzag superconductor wire with smooth edge is also very challenging.

    The working temperature of SSPD ($\sim$3K) is significantly higher than
    that of TES. SSPDs can work in closed-cycle refrigerator.
    Moreover, this relatively high working temperature significantly
    reduces the relaxation time \cite{DPSK:200km}.

    \item[Homodyne Detectors] are used to count the \emph{photon number} of a very weak pulse
    ($\sim100$ photons). The principle is to use a very strong pulse
    (often called the local oscillator) to interfere with the weak
    pulse. Then use two photo diodes to convert the two resulting
    optical pulses into electrical signals, and make a substraction
    between the two electrical signals.

    The homodyne detectors are in general very efficient as there
    is always some detection result given some inputting signals. However,
    the noise of the detectors as well as in the electronics is very
    significant. Moreover, the two photo diodes in the homodyne
    detector have to be identical, which is hard to meet in
    practice.

    The homodyne detectors are commonly used in GMCS QKD systems,
    and they are so far the only choice for GMCS QKD systems.
    Recently, homodyne detectors have also been used to implement the BB84 protocol.

    \end{description}

\subsection{Truly quantum random number generators}\label{se:RNG}

An important but often under-appreciated requirement for QKD is a high data-rate truly quantum
random number generator (RNG). An RNG is needed because most QKD protocols (with the exception
of a passive choice of bases in an entanglement-based QKD protocol) require Alice to choose actively
random bases/signals. Given the high repetition rate of QKD, such a RNG must have a high data-rate.
To achieve unconditional security, a standard software-based pseudo-random number generator
cannot be used because it is actually deterministic. So, a high data rate quantum RNG is a natural
choice. Incidentally, some firms such id Quantique do offer commercial quantum RNGs.
Unfortunately, it is very hard to generate RNG by quantum means at high-speed.
In practice, some imperfections/bias in the numbers generated by a quantum RNG are
inevitable. The theoretical foundation of QKD is at risk because
existing security proofs all assume the existence of perfect RNGs and
do not apply to imperfect RNGs.

\section{Experimental Implementation of BB84 Protocol}

In this section, we will focus mainly on the optical layer.
We will skip several important layers.
In practice, the control/electronics layer is equally important.
Moreover, it is extremely challenging to implement the classical post-processing
layer in real-time, if one chooses block sizes of codes to be long enough to
achieve unconditional security.

\subsection{Optical Layer: Polarization Coding}\label{se:OL_Polarization}

\begin{figure}
  \includegraphics[width=10cm]{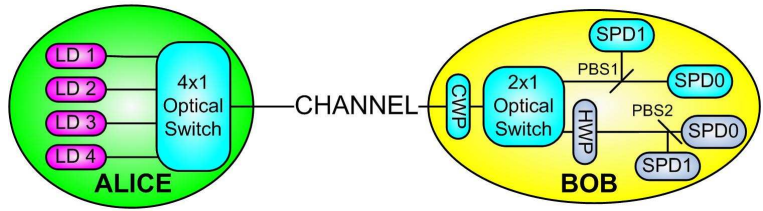}\\
  \caption{Conceptual schematic for polarization-coding BB84 QKD system.
  LD: Laser Diode; CWP: Compensating Wave Plate; HWP: Half Wave Plate; PBS: Polarizing Beam Splitter; SPD: Single Photon Detector.}\label{Fig:PolarizationCoding}
\end{figure}

Polarization coding usually uses four laser sources generating the
four polarization states of BB84 \cite{BB84} protocol. A conceptual
schematic is shown in Figure \ref{Fig:PolarizationCoding}. Note that
due to polarization dispersion of fiber, usually people need some
compensating like the waveplates.

The polarization compensation should be implemented dynamically as
the polarization dispersion in the fiber changes frequently. This is
solved by introducing the electrical polarization controller in
\cite{Decoy:PanPRL}. 


\begin{figure}
  \includegraphics[width=10cm]{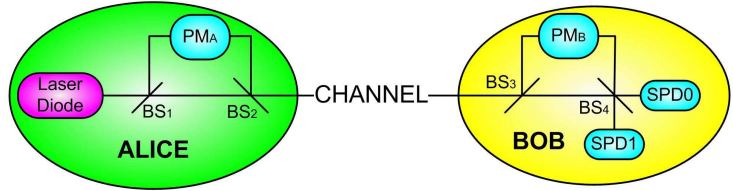}\\
  \caption{Conceptual schematic for double Mach-Zehnder interferometer phase-coding BB84 QKD system.
  PM: Phase Modulator; BS: Beam Splitter; SPD: Single Photon Detector.}\label{Fig:PhaseCoding_DMZ}
\end{figure}

\subsection{Optical Layer: Phase Coding}\label{se:OL_Phase}
\begin{description}
    \item[Original Scheme] is basically a big interferometer as shown in Figure \ref{Fig:PhaseCoding}. 
    However it is not practical as the stability of such a huge
    interferometer is extremely poor.

    \item[Double Mach-Zehnder Interferometer Scheme] is an improved
    version of the original proposal. It has two interferometers and
    there is only one channel connecting Alice and Bob (comparing to the two channels in the original
    proposal). A conceptual set-up is shown in Figure \ref{Fig:PhaseCoding_DMZ}. 
     We can see that the two signals travel through the same
    channel. They only propagate through different paths locally in
    the two Mach-Zehnder interferometers. Therefore people only need
    to compensate the phase drift of the local
    interferometers (the polarization drift in the channel still needs
    to be compensated). This is a great improvement over the original
    proposal. However, the local compensation has to be implemented in real
    time. This is quite challenging. An example set-up that
    implemented the real time compensation of both polarization and
    phase drifting is reported in \cite{Decoy:60Hour}.

\begin{figure}
  \includegraphics[width=10cm]{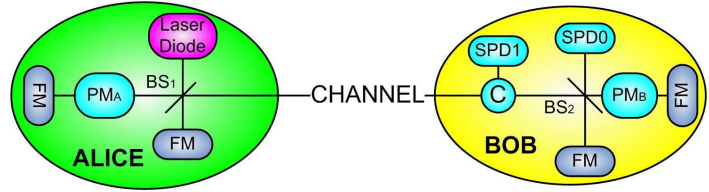}\\
  \caption{Conceptual schematic for Faraday-Michelson phase-coding BB84 QKD system.
  FM: Faraday Mirror, PM: Phase Modulator; BS: Beam Splitter; C: Circulator; SPD: Single Photon Detector.}\label{Fig:PhaseCoding_FM}
\end{figure}

    \item[Faraday-Michelson Scheme] is an improved version of the
    double Mach-Zehnder interferometer scheme. It still has two
    Mach-Zehnder interferometers but each interferometer has only
    one beam splitter. The light propagates through the same section
    of fiber twice due to the Faraday mirror. The schematic is shown
    in Figure \ref{Fig:PhaseCoding_FM}. 
    We can see that the polarization drift is self-compensated. This
    is a great advance in uni-directional QKD implementation and is
    first proposed and implemented in
    \cite{ExpQKD:Faraday_Michelson}.

    Nonetheless, the phase drift of local interferometers still
    needs compensation. Due to the fast fluctuation of phase
    drift (a drift of $2\pi$ usually takes a few seconds), this
    compensation should be done in real-time. A Faraday-Michelson decoy state
    QKD implementation over 123.6km has been reported in
    \cite{Decoy:130km}.

    \begin{figure}
  \includegraphics[width=10cm]{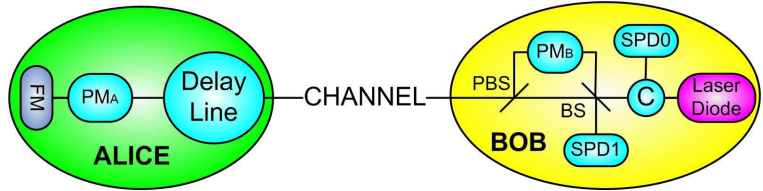}\\
  \caption{Conceptual schematic for ``Plug \& Play'' phase-coding BB84 QKD system.
  FM: Faraday Mirror, PM: Phase Modulator; BS: Beam Splitter; PBS: Polarizing Beam Splitter; C: Circulator; SPD: Single Photon Detector.}\label{Fig:PhaseCoding_PnP}
\end{figure}

    \item[Plug \& Play Scheme] is another improved version of
    the double Mach-Zehnder interferometer scheme. It has only one
    Mach-Zehnder interferometer and the light propagates through the
    same channel and interferometer twice due to the faraday mirror
    on Alice's side. A conceptual set-up is shown in Figure \ref{Fig:PhaseCoding_PnP}. 
    We can see that both the polarization drift and the phase drift
    are automatically compensated. A `Plug \& Play'' scheme based
    decoy state QKD implementation over 60km has been reported in
    \cite{Decoy:ZhaoISIT}.

    Nonetheless, the bi-directional design brings complications to security as
    Eve can make sophisticated operations
    on the bright pulses sent from Bob to Alice. This is often
    called the ``Trojan horse'' attack \cite{Gisin:TrojanHorse}.
    Recently, the security of "Plug and Play" QKD system has been proven in
    \cite{Security:UntrustedSource}.

\begin{figure}
  \includegraphics[width=10cm]{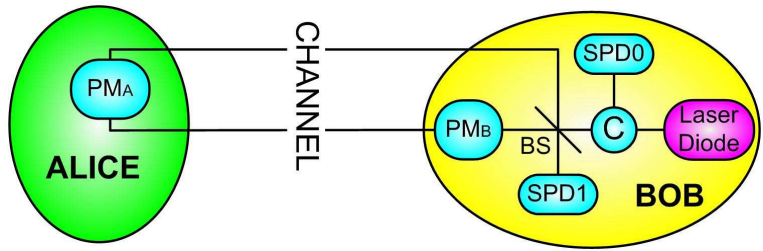}\\
  \caption{Conceptual schematic for Sagnac loop phase-coding BB84 QKD system.
  PM: Phase Modulator; BS: Beam Splitter; C: Circulator; SPD: Single Photon Detector.}\label{Fig:PhaseCoding_Sagnac}
\end{figure}

    \item[Sagnac Loop Scheme.] Another bi-directional optical layer design is to use a Sagnac loop
    where the quantum signal is encoded in the relative phase between the clockwise and
    counter-clockwise pulses that go through the loop. The typical
    schematic is shown in Figure \ref{Fig:PhaseCoding_Sagnac}.

    Sagnac loop QKD is simple to set up and can be easily used in a network setting with a loop
    topology. However, its security analysis is highly non-trivial.

\end{description}
\section{Other Quantum Key Distribution Protocols}

Given the popularity of the BB84 protocol, why should people be
interested in other protocols? There are at least three answers to this question.
First, to better understand the foundations of QKD and its generality, it is
useful to have more than one protocol.
Second, different QKD protocols may have advantages and disadvantages.
They may require different technologies to implement.
Having different protocols allow us to compare and contrast them.
Third, while it is possible to implement standard BB84 protocol with attenuated laser pulses,
its performance in terms of key generation
rate and distance is somewhat limited. Therefore, we have to study other protocols.
Since, from a practical stand point, the third reason is the most important one. We
will elaborate on it in the following paragraph.

The original BB84 \cite{BB84} proposal requires a single photon
        source. However, most QKD implementations are based on faint lasers
        due to the great challenge to build perfect single photon
        sources. Faint laser pulses are weak coherent states that follow
        Poisson distribution for the photon number. The existence of
        multi-photon signals opens up new attacks such as photon-number-splitting attack.
        The basic idea of a photon-number-splitting attack is that
        Eve can introduce a photon-number-dependent transmittance.
        In other words, she can selectively suppress single-photon
signals and transmit multi-photon signals to Bob.
Notice that, for each multi-photon signal, Eve can beamsplit it and keep one copy for herself,
thus allowing her to gain a lot of information about the raw key.

        The security of coherent laser based QKD systems was
        analyzed first against individual attacks in 2000
        \cite{Lutkenhaus:PRA00}, then eventually for a general attack in 2001
        \cite{ILM} and 2002 \cite{GLLP}. Unfortunately, unconditionally secure QKD based on conventional BB84 protocol \cite{ILM,GLLP}
        will severely limit the performance of QKD systems.
        Basically, Alice has to attenuate her source so that the expected
        number $\mu$ of photon is of the same order as the transmittance, $\eta$.
        As a result, the key
        generation rate will scale only quadratically with the transmittance of the channel.

Some of the protocols discussed in the following subsections may
        dramatically improve the performance of QKD over standard BB84 protocol.
        For instance, the decoy state protocol has been proven to provide
        a key generation rate that scales linearly with the transmittance of the channel
        and has been successfully implemented in experiments.

We conclude with some simple alternative QKD protocols. In 1992,
Bennett proposed a protocol (B92) that makes use of only two
non-orthogonal states \cite{B92}. A six-state QKD protocol was first
noted by Bennett and co-workers \cite{SixState} and some years later
by Bruss \cite{SixState_Bruss}. It has an advantage of being
symmetric. Even QKD protocols with orthogonal states have been
proposed \cite{Proposal:OrthogonalStateQKD}. Efficient BB84 and
six-state QKD protocols have been proposed and proven to be secure
by Lo, Chau and Ardehali \cite{Security:Lo_Chau_Ardehali}. A
Singaporean protocol has also been proposed. Recently, Gisin and
co-workers proposed a one-way coherent QKD scheme
\cite{Proposal:Coherent_OneWay}.

    \subsection{Entanglement-based
    Protocols}\label{se:protocol_entanglement}

        \subsubsection{Proposals}

In 1991, Ekert proposed the first entanglement based QKD protocol,
commonly called E91 \cite{Ekert91}. The basic idea is to test the
security of QKD by using the violation of Bell's inequality. Note
that one can also implement the BB84 protocol by using an
entanglement source. Imagine Eve prepares an entangled state of a
pair of qubits and sends one qubit to Alice and the second qubit to
Bob. Each of Alice and Bob randomly chooses one of the two conjugate
bases to perform a measurement.

        \subsubsection{Implementations}
        The key part of entanglement-based quantum cryptography is
        to distribute an entangled pair (usually EPR pair) to two
        distant parties, Alice and Bob.

        Polarization entanglement is preferred in QKD as it is easy
        to measure the polarization (typically via polarizing beam
        splitter). The air has negligible birefringence and thus is
        the perfect channel for polarization-entanglement QKD.

        In free-space QKD, atomspherical turbulence may shift the light beam. Therefore the
        collection of incident photon is challenging. Usually large
        diameter optical telescope is needed to increase the
        collection efficiency.

        \begin{figure}
  \includegraphics[width=10cm]{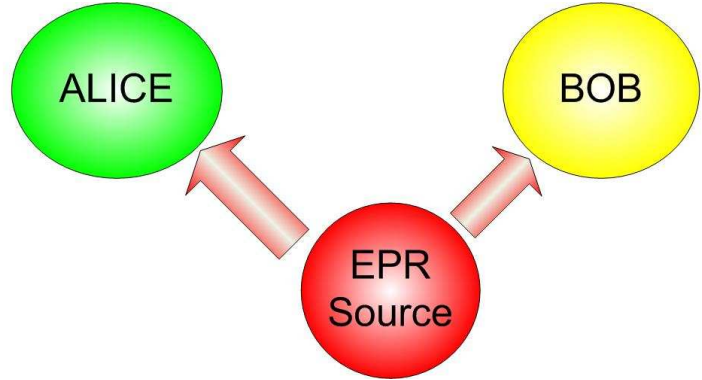}\\
  \caption{Conceptual schematic entanglement-based QKD system with the source in the middle of Alice and Bob.}\label{Fig:EPR_Mid_Source}
\end{figure}

        A standard approach is to
        put the entanglement source right in the middle of Alice and
        Bob. See Figure \ref{Fig:EPR_Mid_Source}. Once an entangled pair is generated, the two particles
        are directed to different destinations. Alice and Bob
        measure the particles locally, and keep the result as the
        bit value. This approach has potential in the ground-satellite
        intercontinental entanglement distribution, in which the entanglement
        source is carried by the satellite and the entangled photons are sent
        to two distant ground stations. A recent source-in-the-middle entanglement-based
        quantum communication work over 13km and  is reported in
        \cite{Entanglement:13km}. 
%

\begin{figure}
  \includegraphics[width=10cm]{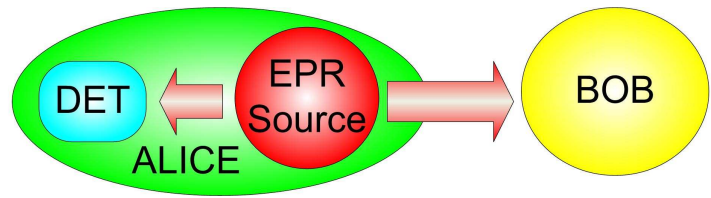}\\
  \caption{Conceptual schematic entanglement-based QKD system with the source at Alice's side. DET: Alice's detection system.}\label{Fig:EPR_Alice_Source}
\end{figure}

         A simpler version is to include the entanglement source in
         Alice's side locally. See Figure \ref{Fig:EPR_Alice_Source}. Once Alice generates an entangled
         pair, she keeps one particle and send the other to Bob.
         Both Alice and Bob measure the particle locally and keep
         the result as the bit value. This approach is significant
         simpler than the above design because only Bob needs the
         telescope and compensating parts. A recent experiment of
         source-in-Alice entanglement-based quantum communication over 144km open air is
         reported in \cite{Entanglement:144km}. 

    \subsection{Decoy State Protocols}\label{se:protocol_decoy}

        \subsubsection{Proposals}

Recall that BB84 implemented with weak coherent state has a key
generation rate that scales only quadratically with the
transmittance. The decoy state protocol can dramatically increase
the key generation rate so that it scales linearly with the
transmittance. In a decoy state protocol, Alice prepares some decoy
states in addition to signal states. The decoy states are the same
as the signal state, except for the expected photon number. For
instance, if the signal state has an average photon number $\mu$ of
order 1 (e.g. $0.5$), the decoy states have an average photon number
$\nu_1$, $\nu_2$, etc. The decoy state idea was first proposed by
Hwang\cite{Decoy:Hwang}, who suggested using a large $\nu$ (e.g $2$)
as a decoy state. Our group provided a rigorous proof of security to
decoy state QKD\cite{Decoy:LoISIT,Decoy:LoPRL}. Our numerical
simulations showed clearly that decoy states provide a dramatic
improvement over non-decoy protocols. In the limit of infinitely
many decoy states, Alice and Bob can effectively limit Eve's attack
to a simple beam-splitting attack. Moreover, we proposed practical
protocols. Instead of using a large $\nu$ as a decoy state, we
proposed using small $\nu$'s as decoy states\cite{Decoy:LoISIT}. For
instance, we proposed using a vacuum state as the decoy state to
test the background and a weak $\nu$ to test the single-photon
contribution. We and Wang analyzed the performance of practical
protocols in detail
\cite{Decoy:Practical,Decoy:WangPRL,Decoy:WangPRA}.

Notice that the decoy state is a rather general idea that can be
applied to other QKD sources. For instance, decoy state protocols
have recently been proposed in
\cite{Decoy:Passive,Decoy:HeraldedSource,Decoy:SimplePDC}  for
parametric down conversion sources. For a comparison of those
protocols, see \cite{Decoy:PDCComparison}.

        \subsubsection{Implementations}

        The first experimental demonstration of decoy state QKD was reported
        by our group in 2006 first over 15 km telecom \cite{Decoy:ZhaoPRL}
        fiber and later over 60 km telecom fiber \cite{Decoy:ZhaoISIT}.
        Subsequently, the decoy state QKD was further demonstrated experimentally by
        several groups worldwide
        \cite{Decoy:144km,Decoy:PanPRL,Decoy:LATES,Decoy:YuanAPL}.

\begin{figure}
  \includegraphics[width=10cm]{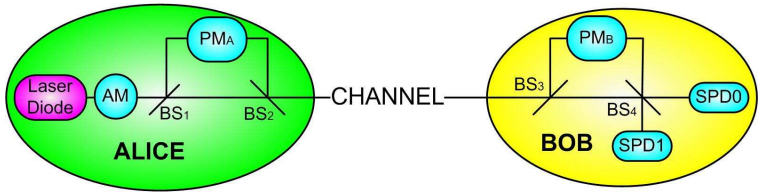}\\
  \caption{Conceptual schematic for decoy state BB84 QKD system (double Mach-Zehnder interferometer
  phase-coding) with amplitude modulator.
  PM: Phase Modulator; AM: Amplitude Modulator; BS: Beam Splitter; SPD: Single Photon Detector.}\label{Fig:Decoy_AM}
\end{figure}

        The implementation of decoy state QKD is straightforward.
        The key part is to prepare signals with different
        intensities. A simple solution is to use an amplitude
        modulator to modulate the intensities of each signal to the
        desired level. See Figure \ref{Fig:Decoy_AM}. Decoy state QKD implementations using
        amplitude modulator to prepare different states are reported
        in
        \cite{Decoy:ZhaoPRL,Decoy:ZhaoISIT,Decoy:YuanAPL,Decoy:LATES,Decoy:144km}.

\begin{figure}
  \includegraphics[width=10cm]{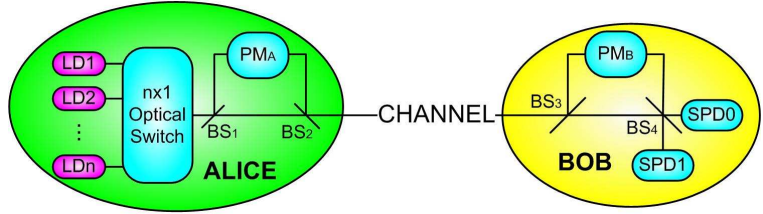}\\
  \caption{Conceptual schematic for decoy state BB84 QKD system (double Mach-Zehnder interferometer phase-coding)
  with multiple laser diodes.
  LDx: Laser Diodes at different intensities; PM: Phase Modulator; BS: Beam Splitter; SPD: Single Photon Detector.}\label{Fig:Decoy_AM}
\end{figure}

        The amplitude modulator has the disadvantage that the
        preparation of vacuum state is quite challenging. An
        alternative solution is to use laser diodes of different
        intensities to generate different states. This solution
        requires multiple laser diodes and high-speed optical
        switch, and is thus more complicated than the amplitude
        modulator solution. Nonetheless, perfect vacuum states can
        be easily prepared in this way. Decoy state QKD implementations using
        multiple laser diodes are reported
        in
        \cite{Decoy:PanPRL,Decoy:130km}.
         Decoy state with a parametric down conversion source has been experimentally implemented
        in \cite{Decoy:ExpPDC_Wang_PRL}.


    \subsection{Strong Reference Pulse
    Protocols}\label{se:protocol_B92}

        \subsubsection{Proposals}

The proposal of strong reference pulse QKD dated back to Bennett's 1992 paper \cite{B92}.
The idea is to add a strong reference pulse, in addition to the signal pulse.
The quantum state is encoded in the relative pulse between the reference pulse
and the signal pulse. Bob decodes by splitting a part of the strong reference pulse
and interfering it with the signal pulse. The strong reference pulse implementation can
counter the photon number splitting attack by Eve because it removes the {\it neutral} signal
in the QKD system. Recall that in the photon number splitting attack, Eve suppresses
single photon signals by sending a vacuum. This works because the vacuum is a neutral signal
that leads to no detection. In contrast, in a strong reference pulse implementation of QKD,
a vacuum signal is not a neutral signal. Indeed,
if Eve replaces the signal pulse by a vacuum and keeps the strong reference pulse unchanged,
then the interference experiment by Bob will give {\it non-zero} detection probability and a random
outcome of "0" or "1". On the other hand, if Eve removes both the signal and the reference pulses,
then Bob may detect Eve's attack by monitering the intensity of the reference pulse,
which is supposed to be strong.

In some recent papers, the unconditional security of B92 QKD with strong reference pulse has been rigorously
proven. However, those proofs require Bob's system to have certain properties and do not apply to standard
threshold detectors.

        \subsubsection{Implementations}
        The B92 \cite{B92} protocol is simpler to implement than
        BB84 \cite{BB84} protocol. However, its weakness in security
        limits people's interest on its implementation. A recent
        implementation of B92 protocol over 200m fiber is reported
        in \cite{ExpB92:200m}. 
%

    \subsection{Gaussian-modulated Coherent State (GMCS)
    Protocol}\label{se:protocol_GMCS}

\begin{figure}
  \includegraphics[width=10cm]{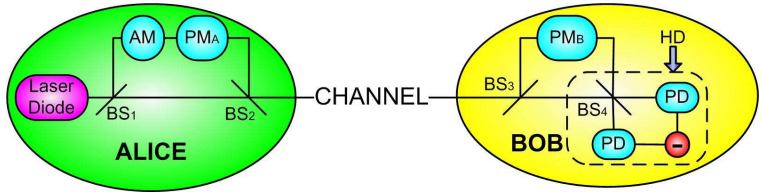}\\
  \caption{Conceptual schematic for Gaussian-modulated Coherent State QKD system.
  PM: Phase Modulator; AM: Amplitude Modulator; BS: Beam Splitter; PD: Photo Diode; HD: Homodyne Detector (inside dashed box).}\label{Fig:GMCS}
\end{figure}

        \subsubsection{Proposals}
Instead of using discrete qubit states as in the BB84 protocol,
one may also use continuous variables for QKD. Early proposals of
continuous variables QKD use squeezed states, which are experimentally challenging.
More recently, gaussian-modulated coherent states have also been proposed for
QKD. Since a laser naturally emits a coherent state, compared to a squeezed
state QKD proposal, a GMCS QKD protocol is experimentally more feasible.
In GMCS QKD, Alice sends Bob a sequence of coherent state signals.
For each signal, Alice draws two random numbers $X_A $ and $P_A$ from a set of
Guassian random number with a mean of zero and a variance of $V_A N_0$
and sends a coherent state $\ket{X_A + i P_A }$ to Bob.
Bob randomly chooses to measure either the $X$ quadrature of the $P$ quadrature
with a phase modulator and a homodyne detector. After performing his measurement, Bob
informs Alice which quadrature he has performed for each pulse, through an
authenticated public classical channel. Alice drops the irrelevant data and keeps only
the quadrature that Bob has measured. Alice and Bob now share a set of correlated Gaussian
variables which they regard as the raw key.
Alice and Bob randomly select a subset of their signals and publicly broadcast their data
to evaluate the excess noise and the transmission efficiency of the quantum channel.
If the excess noise is higher than some prescribed level, they abort.
Otherwise, Alice and Bob perform key generation by
some prescribed protocol.

An advantage of a GMCS QKD is that every signal can be used to generate a key, whereas
in qubit-based QKD such as the BB84 protocol losses can substantially reduce the key generation
rate. Therefore, it is commonly believed that for short-distance (say $< 15$ km) applications,
GMCS QKD may give a higher key generation rate.

GMCS QKD has been proven to be secure only against individual attacks. The security of
GMCS QKD against the most general type of attack---joint attack---remains an open question.

        \subsubsection{Implementations}
            GMCS protocol has significant advantage over the BB84 \cite{BB84} protocol at
            short distances. It was first implemented by F. Grosshans et al. in
            2003 \cite{GMCS:Nature}. It was shown to be working with channel
            loss up to 3.1dB, which is equivalent to the loss of 15km telecom
            fiber. Nonetheless, the strong spectral and polarization dispersion
            of telecom fiber made it challenging to build up a fiber-based GMCS
            system. Lodewyck, Debuisschert, Tualle-Brouri, and Grangier built the first fiber-based GMCS system
            in 2005 \cite{GMCS:OneMeter} but only over a few meters. This
            distance was largely extended to 14km by Legr\'{e}, Zbinden, and Gisin in 2006
            \cite{GMCS:PnP} with the introduction of the ``plug \& play''
            design, which brought questions on its security. The
            uni-directional GMCS QKD has been later implemented over 5km optical
            fiber by Qi, Huang, Qian, and Lo in 2007 \cite{GMCS:5km} and over 25km optical
            fiber by J. Lodewyck et al. in 2007 \cite{GMCS:25km}.

            GMCS QKD requires dual-encoding on both amplitude
            quadrature and phase quadrature, and homodyne detection for
            decoding. See Figure \ref{Fig:GMCS}. Its implementation is in general more challenging than that of BB84 protocol. 

    \subsection{Differential-phase-shift-keying (DPSK)
    Protocols}\label{se:protocol_DPSK}

        \subsubsection{Proposals}

In DPSK protocol, a sequence of weak coherent state pulses is sent from Alice to Bob.
The key bit is encoded in the relative phase of the adjacent pulses.
Therefore, each pulse belongs to two signals.
DPSK protocol also defeats the photon number splitting attack by
removing the neutral signal. Eve may attack a finite train of signals by measuring its
total photon number and then splitting off one photon, whenever the photon number is larger than one.
But, since each pulse belongs to two signals, the pulses in the boundary of the train will
interfere with the pulses immediately outside the boundary.
Therefore, Eve's attack does not allow her to gain full information about
all the bit values associated with the train. Moreover, by splitting the signal,
Eve has reduced the amplitude of the pulses at the boundary of the train. Therefore,
Bob will detect Eve's presence by the higher bit error
rates for the bit values between the pulses at the boundary
and those just outside the boundary.

While DPSK protocol is simpler to implement than BB84, a proof of
its unconditional security is still missing. Therefore, it is hard
to quantify its secure key generation rate and perform a fair
comparison with, for example, decoy state BB84 protocol. Attacks
against DPSK has been studied in, for example,
\cite{ThrHack:Sequencial,ThrHack:SequencialIM}.

\begin{figure}
  \includegraphics[width=10cm]{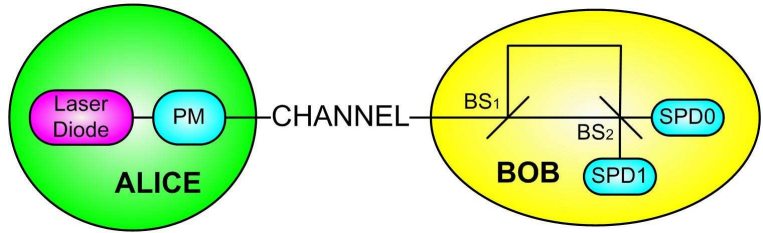}\\
  \caption{Conceptual schematic for differential phase shift keying QKD system.
  PM: Phase Modulator; BS: Beam Splitter; SPD: Single Photon Detector.}\label{Fig:DPSK}
\end{figure}

        \subsubsection{Implementations}
        DPSK protocol is simpler
        in hardware design than the BB84 \cite{BB84} protocol as it requires
        only one Mach-Zehnder interferometer. See Figure \ref{Fig:DPSK}. It also has the potential in
        high-speed applications. Honjo, Inoue, and Takahashi experimentally demonstrated
        this protocol with a planar light-wave circuit over 20km fiber in
        2004 \cite{DPSK:20km}. This distance was soon extended to 105km
        \cite{DPSK:105km} in 2005. In 2007, DPSK scored both the longest and
        the fastest records in QKD implementations: H. Takesue et al.
        reported an experimental demonstration of DPSK-QKD over 200km
        optical fiber at 10GHz \cite{DPSK:200km}. However, since a proof of
        unconditional security is still missing (see last two paragraphs), it is
        unclear whether the existing
        experiments generate any secure key.
        Indeed, the attacks described in \cite{ThrHack:Sequencial,ThrHack:SequencialIM}  showed that with
only one-way classical post-processing, all existing DPSK
experiments are insecure.


\section{Quantum Hacking}
Since practical QKD systems exist and commercial QKD systems are on
the market, it is important to understand how secure they really
are. We remark that there is still a big gap between the theory and
practice of QKD. Even though the unconditional security of practical
QKD systems with semi-realistic models have been proven \cite{ILM,GLLP}, practical
QKD systems may still contain fatal security loopholes. From a
historical standpoint, Bennett and Brassard mentioned that the first
QKD system \cite{EQC} was unconditionally secure to any eavesdropper who
happened to be deaf! This was because the system made different sounds
depending on whether the source was sending a ``0" or a ``1". Just
by listening the sounds, an eavesdropper can learn the value of the
final key. This example highlights the existence of side channels in
QKD and how easy an eavesdropper might be able to break the security
of a QKD system, despite the existence of security proofs.

In this section, we will sketch a few cleverly proposed quantum hacking
strategies that are outside standard security proofs and their
experimental implementations. We will conclude counter-measures and
future outlook. Notice that we will skip eavesdropping attacks that have already been covered by
standard security proofs.

\subsection{Attacks}

1) Large Pulse Attack. In a large pulse attack, Eve sends in a
strong pulse of laser signal to, for example, Alice laboratory to
try to read off Alice's phase modulator setting from a reflected
pulse. See Ref. \cite{ThrHack:TrojanHorse}. As a result, Eve may
learn which BB84 state Alice is sending to Bob.

A simple counter-measure to the large pulse attack is to install an isolator in
Alice's system.

\begin{figure}
  \includegraphics[width=10cm]{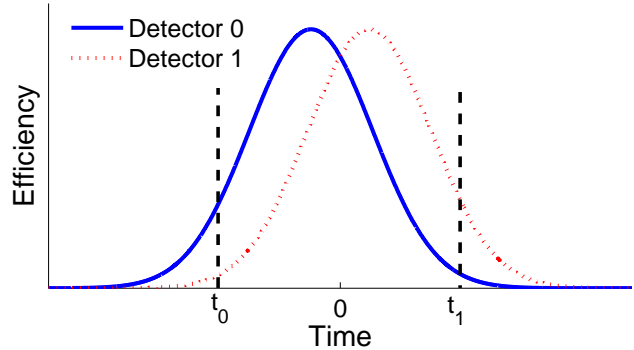}\\
  \caption{Conceptual schematic for detection efficiency mismatch in time domain.}\label{Fig:EfficiencyMismatch}
\end{figure}

2) Faked State Attack. Standard InGaAs detectors suffer from
detection efficiency mismatch. More concretely, as noted in Section
\ref{se:detection}, InGaAs detectors are often operated in a gated
mode. Therefore, the detection efficiency of each detector is
time-dependent. Refer to Figure \ref{Fig:EfficiencyMismatch} for a
schematic diagram of the detection efficiencies of two detectors
(one for ``0" and one for ``1") as functions of time. At the
expected arrival time, the detection efficiency of the two detectors
are similar. However, if the signal is chosen to arrive at some
unexpected times, it is possible that the detector efficiencies of
the two detectors differ greatly.

The faked state attack proposed by Makarov and co-workers
\cite{ThrHack:Makarov_Mismatch} is an intercept-resend attack. In a
faked state attack, for each signal, Eve randomly chooses one of the
two BB84 bases to perform a measurement. Eve then sends to Bob a
{\it wrong} bit in the {\it wrong} basis at a time when the detector
for the wrong bit has a low detection efficiency. For instance, if
Eve has chosen the rectilinear basis and has found a "0" in the bit
value, she then prepares a state "1" in the diagonal basis and sends
it to Bob at the arrival time where detector efficiency of the
detector for "0" is much higher than that of detector for "1".

Now, should Eve have chosen the wrong basis in her measurement,
notice that
the detection probability by Bob is greatly suppressed. For instance, in our example, if the
correct basis is the diagonal basis, let us consider when happens when Bob measures the
signal in the correct basis. Since the bit value resent by Eve is "1" in the diagonal basis
and Bob's detector for "1" has a low detection efficiency, most likely
Bob will not detect any signal.  On the other hand, should Eve have
chosen the correct basis in her measurement, Bob has a significant detection efficiency.
For instance, in our example, if the correct basis is in fact the rectilinear basis, let us
consider what happens when Bob measures in the correct basis.
In this case, a bit "1" in the diagonal basis sent by Eve can be re-written as a superposition of
a bit "0" and a bit "1" in the rectilinear basis. Since the detector for "0" has a much higher
detection efficiency than the detector for "1", most likely Bob will detect a
"0". Since "0" was exactly what was originally sent by Alice, Bob will find a rather low
bit error rate, despite Eve's intercept-resend attack.

The faked state attack, while conceptually interesting, is hard to implement in practice.
This is because it is an intercept-resend attack and as such involves finite detection efficiency
in Eve's detectors and precise synchronization between Eve and Alice-Bob's system.
For this reason, the faked state attack has never been implemented in practice.

3) Time-shift Attack. The time-shift attack was proposed by Qi,
Fung, Lo, and Ma \cite{ThrHack:TimeShift}. It also utilizes the
detection efficiency mismatch in the time domain, but is much easier
to implement than the faked states attack.

As we mentioned in the above section, typical InGaAs-APD detectors
usually operate in a gated mode. That is, if the photon hits the
detector at unexpected time, the two detectors may have
substantially different efficiency. Therefore, Eve can simply shift
the arrival time of each signal, creating large efficiency mismatch
between ``0''s and ``1''s.

Let's take a specific example to illustrate this attack: suppose
detector 0 has higher efficiency than detector 1 if the signal
arrives \emph{earlier} than the expected time, and lower efficiency
than detector 1 if the signal arrives \emph{later} than expected.
Eve can simply shift the arrival time of each bit by sending it
through a longer path or a shorter one. Consider the case in which
Eve sends bit $i$ through a shorter path. In this case bit $i$ will
hit the detector \emph{earlier} than expected, thus detector 0 has
much higher efficiency. If Bob reports a detection event for the
$i$th bit, Eve can make a guess that this bit is a ``0'' with high
probability of success.

Furthermore, Eve can carefully set how many bits should be shifted
forward and how many should be shifted backward so that Bob gets
similar counts of ``0''s and ``1''s. In this way, Bob cannot observe
a mismatch between the numbers of ``0''s and ``1''s.

Note that the time-shift attack does not make any measurement on the
qubits. Therefore, quantum information is not destroyed. That is,
Eve does not change the polarization, the phase, or the frequency of
any bit. This means the time-shift attack will not increase the bit error rate
of the system in principle. Moreover, since Eve does not need to
make any measurement or state preparation, the time-shift attack is
practically feasible even with current technology.

The time-shift attack will introduce some loss as the overall
detection efficiency is lower if the photon hits the detector at an
unexpected time. Nonetheless, Eve can compensate this loss by making
the channel more transparent. Notice that, since the quantum channel between Alice and Bob
may contain many lossy components such as splices and couplers, it may
not be too hard for Eve to make a channel more transparent.

The time-shift attack has been successfully implemented on a
commercial QKD system by Zhao, Fung, Qi, Chen, and Lo
\cite{ExpHack:TimeShift} in 2007. This is the first and so far the
only experimentally successful demonstration of quantum hacking on
commercial QKD system. It is shown that the system has no-negligible
probability to be vulnerable to the time-shift attack. Quantitative
analysis shows that the final key shared by Alice and Bob (after the
error correction and the privacy amplification of the most general
security analysis) has been compromised by Eve.

The success of the time-shift attack in \cite{ExpHack:TimeShift} is
rather surprising as QKD has been widely believed to be
unconditionally secure. The experimental success in quantum hacking
highlighted the limit of the whole research program of
device-independent security proofs \cite{Security:DeviceIndependent}
by showing that device-independent security proofs, even if they are
found to exist in future, do not apply to a practical QKD system.
The success of time-shift attack is not due to some technical
imperfection. It is deeply connected with the detection efficiency
loophole in the verification of Bell-inequalities. So far the InGaAs
detectors have only $\sim10\%$ detection efficiencies, and the
channel connecting Alice and Bob usually has quite large attenuation
for long distance communication. The low overall detection
efficiency fails the device-independent security proof.

Notice that even non-gated detectors have dead times and generally
suffer from detection efficiency loophole. The detection efficiency
mismatch is also discussed in
\cite{ThrHack:TrojanHorseBellInequality}.

4) Phase remapping attack. In a bi-directional implementation of QKD
such as the "Plug and Play" set-up, Eve may attempt to tamper with
Alice's preparation process so that Alice prepares four wrong
states, instead of the four standard BB84 state. This is called the
phase remapping attack and was proposed in
\cite{ThrHack:PhaseRemapping}.

In a "Plug and Play" QKD system, Alice receives a strong pulse from Bob and she
then attenuates it to a single-photon level and encodes one of the BB84 state on it.
For instance, Alice may encode her state by using a phase modulator.
In a phase modulator, the encoded phase is proportional to the voltage applied.
In practice, a phase modulator has a finite rise time. For each BB84 setting,
one may thus model the applied
voltage (and thus the encoded phase) as a trapezium. Ideally, Bob's strong
pulse should arrive at the plateau region of the phase
modulation, thus getting maximal phase modulation by
Alice's phase modulator.
Now, imagine that Eve applies a time-shift to Bob's strong pulse so that
it arrives in the rise region of the phase modulation graph instead of the plateau region.
In this case, Alice has wrongly encoded her phase only partially.

If we assume that the four settings of Alice's encoding (for the four BB84 states) have
the same rise region, then by time-shifting Bob's strong pulse, Eve can force Alice to
prepare the four state with phase $0, a, 2a, 3a$, rather than $0, \pi/2, \pi, 3 \pi/2$.
In general, these four states are more distinguishable than the standard BB84 state.
Therefore, Eve may subsequently apply an intercept-resend attack to the signal sent out by Alice.

It was proven in \cite{ThrHack:PhaseRemapping} that in principle Eve
can break the security of the QKD system, without alerting Alice and
Bob.

5) Attack by passive listening to side channels. The attack by
listening to the sounds made by the source in the first QKD
experiment is an example of an attack by passive listening to side
channels. Another example is \cite{ExpHack:DetectionTime}. 
A counter-measure is to carefully locate all possible side channels and to
eliminate them one by one.

6) Saturation Attack: In a recent preprint
\cite{ThrHack:Makarov_Sacturation}, Makarov studied experimentally
how by sending a moderately bright pulse, Eve can blind Bob's InGaAs
detector. A simple counter-measure would be for Bob to measure the
intensity of the incoming signal.

7) High Power Damage Attack: In Makarov's thesis, it was proposed
that Eve may try to make controlled changes in Alice's and Bob's
system by using high power laser damage through sending a very
strong laser pulse. Again, a simple counter-measure would be for
Alice and Bob to measure the intensity of the incoming signals and
monitor the properties of various components from time to time to
ensure that they perform properly.

\subsection{Counter-measures}
Once an attack is known, there are often simple counter-measures.
For instance, for the large pulse attack, a simple counter-measure
would be to add a circulator in Alice's laboratory. As for the faked
state attack and time-shift attack, a simple counter-measure would
be for Bob to use a four-state setting in his phase modulator. Other
counter-measures include Bob applying a random time-shift to his
received signals. However, the most dangerous attacks are the {\it
unanticipated} ones.

Notice that it is not enough to say that a counter-measure to an
attack exists. It is necessary to actually implement a
counter-measure experimentally in order to see how effective and
convenient it really is. This will allow Alice and Bob to select a
useful counter-measure. Moreover, notice that the implementation of
a counter-measure may itself open up new loopholes. For instance, if
Bob implements a four-state setting as a counter-measure to a
time-shift attack, Eve may still combine a large pulse attack with
the time-shift attack to break a QKD system.

\subsection{Importance of quantum hacking}

As noted in Section \ref{se:security_proof}, there has been a lot of
theoretical interest on the connection between the security of QKD
and fundamental physical principles such as the violation of Bell's
inequality. An ultimate goal of such investigations, which has not been realized yet, is to construct
a device-independent security proof
\cite{Security:DeviceIndependent}. Even if such a goal is achieved in future,
would any of these theoretical
security proofs applies to a quantum key distribution system in {\it
practice}? Unfortunately, the answer is no. As is well-known, the
experimental testing of Bell-inequalities often suffers from the
detection efficiency loophole \cite{Security:DeviceIndependent}. The
low detection efficiency of practical detectors not only nullifies
security proofs based on Bell-inequality violation, but also gives
an eavesdropper a powerful handle to break the security of a
practical QKD system. Therefore, the detection efficiency loophole
is of both theoretical and practical interest.

A practical QKD system often consists of two or more detectors. In
practice, it is very hard to construct detectors of identical
characteristics. As a result, two detectors can generally exhibit
different detector efficiencies as functions of either one or a
combination of variables in the time, frequency, polarization or
spatial domains. Now, if an eavesdropper could manipulate a signal
in these variables, then she could effectively exploit the detection
efficiency loophole to break the security of a QKD system. In fact,
she could even violate a Bell-inequality with only a classical
source. In time-shift attack, one consider an eavesdropper's
manipulation of the time variable. However, the generality of
detection efficiency loophole and detector efficiency mismatch
should not be lost.

We should remark that, for eavesdropping attacks, the sky is the limit.
The more imaginative one is, the more new attacks one comes up.
Indeed, what people have done so far are just scratching the surface of the subject.
Much more work needs to be done in the battle-testing of QKD systems and security proofs
with testable assumptions. See Section on Future Directions.

\section{Beyond Quantum Key Distribution}

Besides QKD, many other applications of quantum cryptography have
been proposed. Consider, for instance, the millionaires' problem.
Two millionaires, Alice and Bob, would like to determine who is
richer without disclosing the actual amount of money each has to
each other. More generally, in a secure two-party computation, two
distant parties, Alice and Bob, with private inputs, $x$ and $y$
respectively, would like to compute a prescribed function $f(x, y)$
in such a way that at the end, they learn the outcome $f(x,y)$, but
nothing about the other party's input, other than what can be
logically be deduced from the value of $f(x,y)$ and his/her input.
There are many possible functions $f(x,y)$. Instead of implementing
them one by one, it is useful to construct some cryptographic
primitives, which if available, can be used to implement the secure
computation of {\it any} function $f(x,y)$. In classical
cryptography, to implement secure two-party computation of a general
function will require making additional assumptions such as a
trusted third party or computational assumptions. The question is
whether we can do unconditionally secure {\it quantum} secure
two-party computations.

Two important cryptographic primitives are namely quantum bit
commitment (QBC) and one-out-of-two quantum oblivious transfer
(QOT). In particular, it was shown by Kilian
\cite{C.Cryp:Kilian} that in classical cryptography, oblivious
transfer can be used to implement a general two-party secure
computation of any function $f(x,y)$. Moreover, in quantum
cryptography, it was proven by Yao \cite{QBC_QOT:Yao} that a secure
QBC scheme can be used to implement QOT securely. For a long time
back in the early 1990's, there was high hope that QBC and QOT could
be done with unconditional security. In fact, in a paper
\cite{QBC:BCJL} it was claimed that QBC can be made unconditionally
secure. The sky fell around 1996 when Mayers \cite{QBC:Mayers} and
subsequently, Lo and Chau \cite{QBC:LoChau}, proved that, contrary
to widespread belief at that time, unconditionally secure QBC is, in
fact, impossible. Subsequently, Lo \cite{QBC:Lo} proved explicitly
that unconditionally secure one-out-of-two QOT is also impossible.
Mayers and Lo-Chau's result was a big step backwards and thus a big
disappointment for quantum security.

After the fall of QBC and QOT, people turned their attention to
quantum coin tossing (QCT). Suppose Alice and Bob are having a
divorce and they would like to determine by a coin toss who is going
to keep their kid. They do not trust each other.
However, they live far away from each other and
have to do a coin toss remotely. How can they do so without trusting
each other? Classically, coin tossing will require either a trusted
third party or making computational assumptions. As shown by Lo and
Chau, ideal quantum coin tossing is impossible
\cite{CoinTossing:LoChau}. Even for the non-ideal case,
Kitaev has proven that a strong version of QCT
(called {\it strong} QCT) cannot be unconditionally secure.
However, despite numerous papers on the
subject (See, for example, \cite{CoinTossing:Mochon} and references therein), whether
non-ideal {\it weak} QCT is possible remains an open question.

Other QKD protocols are also of interest. For instance, the sharing
of quantum secrets has been proposed in
\cite{Proposal:SecretSharing_CleveGottesmanLo}. It is an important
primitive for building other protocols such as secure multi-party
quantum computation \cite{Proposal:SecureMultiQuantumComputation}.
There are also protocols for quantum digital signatures
\cite{Proposal:QuantumDigitalSignature}, quantum fingerprinting and
unclonable encryption. Incidentally, quantum mechanics can also be
used for the quantum sharing of classical secrets
\cite{Proposal:QuantumSecretSharing}, conference key agreement and
third-man cryptography.

For QKD, so far we have only discussed a point-to-point
configuration. In real-life applications, it will be interesting to
study QKD in a network setting \cite{SECOQC:WhitePaper}.  Note that
the multiplexing of several QKD channels in the single fiber has
been successful performed. So has the multiplexing of a classical
channel together with a QKD channel. However, much work remains to
be done on the design of both the key management structure and the
optical layer of a QKD network.

\section{Future Directions}

The subject of quantum cryptography is still in a state of flux. We will conclude with
a few examples of future directions.

\subsection{Quantum Repeaters}
Losses in quantum channels greatly limit the distance and key
generation rate of QKD. To achieve secure QKD over long distances
without trusting the intermediate nodes, it is highly desirable to
have quantum repeaters. Briefly stated, quantum repeaters are
primitive quantum computers can be perform some form of quantum
error correction, thus preserving the quantum signals used in QKD.
In more detail, quantum repeaters often rely on the concept of entanglement distillation,
whose goal is, given a large number $M$ of noisy entangled states, two parties, Alice and Bob,
perform local operations and classical communications to distill out a smaller number (say $N$)
but less noisy entangled states.

The experimental development of a quantum repeater will probably involve the development
of quantum memories together with the interface between flying qubits and qubits in a quantum memory.

\subsection{Ground to satellite QKD}
Another method to extend the distance of QKD is to perform QKD
between a satellite and a ground station. If one trusts a satellite,
one can even build a global QKD network via a satellite relay.
Basically, a satellite can perform QKD with Alice first, when it has
a line of sight with Alice. Afterwards, it moves in orbit until it
has a line of sight with Bob. Then, the satellite performs a
separate QKD with Bob. By broadcasting the XOR of the two keys,
Alice and Bob will share the same key. Satellite to ground QKD
appears to be feasible with current or near-future technology, for a
discussion, see, for example, \cite{Proposal:GroundSatellite}.

With an untrusted satellite, one can still achieve secure QKD between two
ground stations by putting an entangled source at the satellite and sending one
half of each entangled pair to each of Alice and Bob.

\subsection{Calculation of the quantum key capacity}

Given a specific theoretical model, so far it is not known how to calculate
the actual secure key generation rate in a noisy channel.
All is known is how to calculate some upper bounds and lower bounds.
This is a highly unsatisfactory situation because we do not really know
the actual fundamental limit of the system.
Our ignorance can be highlighted by a simple open question:
what is the highest tolerable bit error rate of BB84 that will still allow
the generation of a secure key?

While lower bounds are known \cite{Security:GottesmanLo,Security:RennerGisinKraus,Security:Chau}
and 25 percent is an upper bound set by a simple intercept-resend attack, we
do not know the answer to this simple question.

Notice that this question is of both fundamental and practical interests.
Without knowing the fundamental limit of the key generation rate, we do not
know what the most efficient procedure for generating a key in a practical setting is.

\subsection{Multi-party quantum key distribution and entanglement}

Besides its technological interest, QKD is of fundamental interest
because it is deeply related to the theory of entanglement, which is
the essence of quantum mechanics. So far there have been limited
studies on multi-party QKD. Notice that there are many deep
unresolved problems in {\it multi}-party entanglement. It would be
interesting to study more deeply multi-party QKD and understand
better its connection to multi-party entanglement. Hopefully, this
will shed some light on the mysterious nature of multi-party
entanglement.

\subsection{Security proofs with testable assumptions}
The surprising success of quantum hacking highlights the big gap
between the theory and practice of QKD. In our opinion, it is
important to work on security proofs with {\it testable}
assumptions. Every assumption in a security proof should be written
down and experimentally verified. This is a long-term research
program.

\subsection{Battle-testing QKD systems} Only through battle-testing can we
gain confidence about the security of a real-life QKD system.
Traditionally, breaking a cryptographic systems is as important as building one.
Therefore, we need to re-double our efforts on the study of eavesdropping attacks
and their counter-measures.

As stated before, quantum cryptography enjoys forward security. Thanks to
the quantum no-cloning theorem, an eavesdropper Eve does not have a transcript of
all quantum signals sent by Alice to Bob. Therefore, once a QKD process has been
performed, the information is gone and it will be too late for Eve to go back to
eavesdrop. Therefore, for Eve to break a real-life QKD system today, it is imperative for Eve to
invest in technologies for eavedropping now, rather than in future.

\section{Acknowledgment}

We thank various funding agencies including NSERC, CRC program, QuantumWorks, CIFAR, MITACS, CIPI, PREA,
CFI,
and OIT for their financial support.

\section{Secondary References}

1) D. Gottesman and H.-K. Lo, "From Quantum Cheating to Quantum Security",
Physics Today, Nov. 2000, p. 22. On-line Available at
http://www.aip.org/pt/vol-53/iss-11/p22.html

2) G. Brassard, "A Bibliography of Quantum Cryptography", http://www.cs.mcgill.ca/~crepeau/CRYPTO/Biblio-QC.html

3) N. Gisin, G. Ribordy, W. Tittel and H. Zbinden, "Quantum Cryptography",
Rev. Mod. Phys. 74, 145 - 195 (2002). On-line Available at http://arxiv.org/abs/quant-ph/0101098

4) H.-K. Lo and N. Lutkenhaus, "Quantum Cryptography: from theory to practice"
Invited paper for "Physics In Canada", Sept.-Dec. 2007. On-line available at
http://arxiv.org/abs/quant-ph/0702202

5) V. Scarani, H. Bechmann-Pasquinucci, N. J. Cerf, M. Dusek, N. Lutkenhaus, andp M. Peev, "A Framework for Practical Quantum Cryptography", On-line Available at http://arxiv.org/abs/0802.4155

6) N. Gisin and R. Thew, "Quantum Communication", Nature Photonics, vol 1, No. 3, pp165-171 (2007).
On-line available at http://arxiv.org/abs/quant-ph/0703255

7) Wikipedia, "Quantum Cryptography", http://en.wikipedia.org/wiki/Quantum\_cryptography

%
%
%
%
%
%
%
%
%
%
%
%
%
%
%
%
%

\bibliography{Z:/reference}

\begin{thebibliography}{115}
\bibitem[{von()}]{von_Neumann_entropy}
\bibinfo{note}{Given a density matrix, $\rho$, define the von Neumann entropy
  of $\rho$ as $$ S(\rho)=-\sum_i\lambda_i\log_2\lambda_i=-{\rm
  tr}\rho\log_2\rho $$ where $\lambda_i$'s are eigenvalues of the density
  matrix $\rho$.}
\end{thebibliography}

\bibliographystyle{apsrev}
\end{document}